\documentclass[namedreferences, hyperref]{solarphysics}
\usepackage[hyperref,optionalrh,solaromanenum]{spr-sola-addons} 
\usepackage{graphicx}                    
\usepackage{color}                       
\usepackage{breakurl}                         

\begin{document}

\newcommand{\ie}{{\it i.e. }}
\newcommand{\eg}{{\it e.g. }}

\newcommand{\adv}{    {\it Adv. Space Res.}} 
\newcommand{\annG}{   {\it Ann. Geophys.}} 
\newcommand{\aap}{    {\it Astron. Astrophys.}}
\newcommand{\aaps}{   {\it Astron. Astrophys. Suppl.}}
\newcommand{\aapr}{   {\it Astron. Astrophys. Rev.}}
\newcommand{\ag}{     {\it Ann. Geophys.}}
\newcommand{\aj}{     {\it Astron. J.}} 
\newcommand{\apj}{    {\it Astrophys. J. Suppl. Ser.}}
\newcommand{\apjs}{    {\it Astrophys. J.}}
\newcommand{\apjl}{   {\it Astrophys. J. Lett.}}
\newcommand{\apss}{   {\it Astrophys. Space Sci.}} 
\newcommand{\cjaa}{   {\it Chin. J. Astron. Astrophys.}} 
\newcommand{\gafd}{   {\it Geophys. Astrophys. Fluid Dyn.}}
\newcommand{\grl}{    {\it Geophys. Res. Lett.}}
\newcommand{\ijga}{   {\it Int. J. Geomagn. Aeron.}}
\newcommand{\jastp}{  {\it J. Atmos. Solar-Terr. Phys.}} 
\newcommand{\jgr}{    {\it J. Geophys. Res.}}
\newcommand{\mnras}{  {\it Mon. Not. Roy. Astron. Soc.}}
\newcommand{\nat}{    {\it Nature}}
\newcommand{\pasp}{   {\it Pub. Astron. Soc. Pac.}}
\newcommand{\pasj}{   {\it Pub. Astron. Soc. Japan}}
\newcommand{\pre}{    {\it Phys. Rev. E}}
\newcommand{\solphys}{{\it Solar Phys.}}
\newcommand{\sovast}{ {\it Soviet  Astron.}} 
\newcommand{\ssr}{    {\it Space Sci. Rev.}} 
\chardef\us=`\_

\begin{article}

\begin{opening}

\title{Probabilistic Drag-Based Ensemble Model (DBEM) Evaluation for Heliospheric Propagation of CMEs}

%
\author[addressref={aff1},corref,email={jcalogovic@geof.hr}]{\inits{J.C.}\fnm{Ja\v{s}a}~\lnm{\v{C}alogovi\'{c}}\orcid{0000-0002-4066-726X}}
\author[addressref={aff1}]{\inits{M.D.}\fnm{Mateja}~\lnm{Dumbovi\'{c}}\orcid{0000-0002-8680-8267}}
\author[addressref={aff1}]{\inits{D.S.}\fnm{Davor}~\lnm{Sudar}\orcid{0000-0002-1196-6340}}
\author[addressref={aff1}]{\inits{B.V.}\fnm{Bojan}~\lnm{Vr\v{s}nak}\orcid{0000-0002-0248-4681}}
\author[addressref={aff1}]{\inits{K.M.}\fnm{Karmen}~\lnm{Martini\'{c}}\orcid{0000-0002-9866-0458}}
\author[addressref={aff2}]{\inits{M.T.}\fnm{Manuela}~\lnm{Temmer}\orcid{0000-0003-4867-7558}}
\author[addressref={aff2,aff3}]{\inits{A.V.}\fnm{Astrid}~\lnm{Veronig}\orcid{0000-0003-2073-002X}}

\runningauthor{J. \v{C}alogovi\'{c} et al.}
\runningtitle{\textit{Solar Physics} Drag-Based Ensemble Model}

\address[id={aff1}]{Hvar Observatory, Faculty of Geodesy, University of Zagreb, Ka\v{c}i\'{c}eva 26, HR-10000 Zagreb, Croatia}
\address[id={aff2}]{Institute of Physics, University of Graz, Austria}
\address[id={aff3}]{Kanzelh\"{o}he Observatory for Solar and Environmental Research, University of Graz, Austria}

\begin{abstract}
The Drag-based Model (DBM) is a 2D analytical model for heliospheric propagation of Coronal Mass Ejections (CMEs) in ecliptic plane predicting the CME arrival time and speed at Earth or any other given target in the solar system. It is based on the equation of motion and depends on initial CME parameters, background solar wind speed, $w$ and the drag parameter $\gamma$. A very short computational time of DBM ($<$ 0.01s) allowed us to develop the Drag-Based Ensemble Model (DBEM) that takes into account the variability of model input parameters by making an ensemble of n different input parameters to calculate the distribution and significance of the DBM results. Thus the DBEM is able to calculate the most likely CME arrival times and speeds, quantify the prediction uncertainties and determine the confidence intervals. A new DBEMv3 version is described in detail and evaluated for the first time determing the DBEMv3 performance and errors by using various CME-ICME lists as well as it is compared with previous DBEM versions. The analysis to find the optimal drag parameter $\gamma$ and ambient solar wind speed $w$ showed that somewhat higher values ($\gamma \approx 0.3 \times 10^{-7}$ km$^{-1}$, $w \approx$ 425 km\,s$^{-1}$) for both of these DBEM input parameters should be used for the evaluation compared to the previously employed ones. Based on the evaluation performed for 146 CME-ICME pairs, the DBEMv3 performance with mean error (ME) of -11.3 h, mean absolute error (MAE) of 17.3 h was obtained. There is a clear bias towards the negative prediction errors where the fast CMEs are predicted to arrive too early, probably due to the model physical limitations and input errors (\eg CME launch speed). This can be partially reduced by using larger values for $\gamma$ resulting in smaller prediction errors (ME = -3.9 h, MAE = 14.5 h) but at the cost of larger prediction errors for single fast CMEs as well as larger CME arrival speed prediction errors. DBEMv3 showed also slight improvement in the performance for all calculated output parameters compared to the previous DBEM versions. 
\end{abstract}

%

\end{opening}

%
\section{Introduction}
\label{s:introduction} 

To understand space weather driven by solar activity is important as it can affect various human technologies, health as well as it can have major implications for the space environment near the Earth and the Earth's atmosphere \citep{Boteler-1998, Schrijver-Mitchell-2013, Cid-2014, Schrijver-2015, Koskinen-2017}. Coronal mass ejections (CMEs) are major drivers of strong geomagnetic storms \citep{Gosling-1991, Richardson-2001, Srivastava-2004, Koskinen-Huttunen-2006, Zhang-2007, Richardson-Cane-2012, Kilpua-2017} and CME-driven shocks have also a dominant role in generating large solar energetic particle (SEP) events \citep{Gopalswamy-2003, Dierckxsens-2015}. Therefore, in order to better support space weather forecasts various propagation models have been developed to predict the CME arrival time and speed as well as other important space weather forecast parameters.

There are several types of CME/shock propagation models \citep{Zhao-dryer-2014, Vourlidas-2019, Verbeke-2019} that can be roughly divided into: simple empirical models \citep{Brueckner-1998, Gopalswamy-2001, Paouris-Mavromichalaki-2017}, machine learning models \citep{Sudar-2016}, drag-based models \citep{Vrsnak-2013, Vrsnak-2014, Zic-2015, Rollett-2016, Napoletano-2018, Dumbovic-2021}, physics-based models \citep{Smart-Shea-1985, Fry-2001, Takahashi-2017} and more complex numerical MHD models \citep{Odstrcil-2004, Detman-2006, Pizzo-2011, Pomoell-Poedts-2018}. Despite the different input, approach, assumptions and complexity, various models show similar performance to predict the CME/shock arrival times within an average error $\pm$10 h and standard deviations often exceeding 20 h \citep{Riley-2018, Vourlidas-2019}.

The drag-based models are based on the equation of motion of CMEs that is determined by the MHD drag force from the background solar wind acting on the CME \citep{Cargill-1996, Gopalswamy-2000, Owens-Cargill-2004}. Numerous observational studies showed that CMEs slower than the solar wind accelerate whereas CMEs faster than the solar wind decelerate \citep{Vrsnak-2001, Vrsnak-2004, Vrsnak-2007, Temmer-2011, Hess-Zhang-2014}. Thus the simple kinematical drag-based model (DBM) for CME propagation was established  \citep{Vrsnak-2007}. 

The CME propagation in DBM is based on CME initial properties: CME launch date and time ($t_0$), CME launch speed ($v_0$), CME starting radial distance ($R_0$), the empirically determined drag parameter ($\gamma$) as well as the properties of the ambient solar wind or specifically background solar wind speed ($w$). Additionally, DBM uses a 2D flattening cone CME geometry that includes the CME’s angular half-width ($\lambda$) and longitude of the CME source region ($\phi_{\mathrm{CME}}$) as additional input parameters \citep{Zic-2015}. If we assume that the drag parameter and solar wind speed are constant ($\gamma$ = const. and $w$ = const.) the equation of CME motion (see \eg \citealt{Vrsnak-2013}) has an analytical solution and the drag acceleration/deceleration has a quadratic dependence on the relative speed between CMEs and the background solar wind \citep{Vrsnak-2007, Vrsnak-2013}. 

Comparison with other propagation models such as the more complex numerical WSA-ENLIL+Cone model \citep{Odstrcil-2004, Wold-2018} showed that DBM gives a similar overall performance and comparable prediction errors \citep{Vrsnak-2014}. Due to its simplicity and very fast computational time, DBM was implemented as an online tool for space weather forecast at several places: i) the Space Situational Awareness (SSA) European Space Agency (ESA) portal\footnote{See \url{http://swe.ssa.esa.int/heliospheric-weather}.}, ii) COMESEP alert system\footnote{See \url{http://comesep.aeronomy.be}.} \citep{Dumbovic-2017}, iii) as one of the given models at Community Coordinated Modeling Center (CCMC)\footnote{See \url{https://ccmc.gsfc.nasa.gov}.} and iv) the Hvar Observatory website\footnote{See \url{http://oh.geof.unzg.hr}.}.

However, the main problem of models are limited and unreliable observations that are needed for the input. This can introduce a large errors in calculation of the CME arrival time ($-1.7 \pm 18.3$ h; Vršnak et al., 2014) and speed when DBM forecasts and observations are compared. The main advantage of analytical DBM is its very fast computational time ($<$  0.01s) allowing to use an ensemble modelling approach and to provide a probabilistic forecasting of CME arrival time and speed in real-time (in less than a minute) for what would be needed several hours in the case of numerical models (\eg ENLIL). The Drag-Based Ensemble Model (DBEM) considers the variability of model input parameters (\eg errors and uncertainties in CME measurements) by making an ensemble of $n$ different input parameters to calculate a distribution and significance of the DBM results. With this approach, DBEM can calculate most likely CME arrival times and speeds, quantify the prediction uncertainties and determine the forecast confidence intervals. Such ensemble approach was also recently implemented in other models such as ENLIL \citep{Mays-2015}. \citet{Dumbovic-2018} evaluated DBEM using the same sample as \citet{Mays-2015} and found comparable model errors as in the case of ENLIL with mean error of -9.7 h and mean absolute error of 14.3 h. Recently, DBEMv3 (version 3) was also compared with the empirical Effective Acceleration Model, EAMv3 \citep{Paouris-2021}. Detailed description of various DBEM versions and their differences is given in the next Section \ref{DBEM-description}.

In this study we use three different CME-ICME pair samples to evaluate, validate and determine the performance of the recently developed DBEMv3 for the first time. The first sample of 16 CME-ICME pairs in the period from March 2013 to June 2014 is taken from \citet{Dumbovic-2018} and it is used to validate and compare DBEMv3 with previous versions of DBEM (Section \ref{DBEM-comparison}). The second CME-ICME sample (9 events) is used to further extend the first sample to April 2011 (see Table \ref{table1-CME-list}). To get the reliable statistics and model performance during longer period of time DBEMv3 is also evaluated using the third larger CME-ICME pairs sample based on the more comprehensive Richardson \& Cane ICME list \citep{Richardson-Cane-2010} consisting of altogether 146 CME-ICME pairs in the period from December 1996 to December 2015 (Section \ref{RC-list-analysis}). Here the DBEMv3 performance is tested with two different model input setups for drag parameter ($\gamma$) and background solar wind speed ($w$) based on the results obtained with the reverse modelling. In first case $\gamma$ value is dependent on the CME launch speed (variable $\gamma$) and in the second case, $\gamma$ is held as fixed value. The so-called reverse modelling with DBEMv3 is performed using the first sample in order to find the optimal model input for the drag parameter $\gamma$ and solar wind speed $w$ (Section \ref{DBEM-gamma-w}).

\section{DBEM Description and Versions}\label{DBEM-description}

The Drag-based Ensemble Model (DBEM) provides probabilistic predictions of the propagation of CMEs in the ecliptic plane and estimates the CME arrival time, speed and acceleration at Earth or any other target in the heliosphere (planets and spacecraft). It is based on the analytical Drag-based Model (DBM) assuming the propagation of CMEs in interplanetary space is only influenced by aerodynamic drag force and the CME propagation is determined by the initial CME properties and the ambient solar wind \citep{Vrsnak-2010, Vrsnak-2013, Dumbovic-2021}. The assumed 2D geometry for DBM is a cone and the CME leading edge is initially a semicircle defined by the CME angular width that flattens with time \citep{Zic-2015, Dumbovic-2021}. In order to have the analytical solution in DBM, the assumption should be made that solar wind speed ($w$) and drag parameter ($\gamma$) are constant. In general, this should be valid for radial distances beyond 15 R$_{\odot}$ assuming that the CME moves in an isotropic constant solar wind and the ambient density has the same fall-off rate as the CME expansion \citep{Vrsnak-2013, Zic-2015}. Due to the Lorentz force that accelerates the CME, the drag force may not be the dominant force at radial distances below 15 R$_{\odot}$ (solar radii) and the assumption about constant $w$ and $\gamma$ may not be valid \citep{Vrsnak-2001, Vrsnak-2004, Sachdeva-2015, Sachdeva-2017}. The DBM assumption is typically valid beyond 15-20 R$_{\odot}$, but for the very slow CMEs, the Lorentz force may be relevant at larger radial distances \citep{Gopalswamy-2000, Gopalswamy-2001, Reiner-2007, Temmer-2011, Sachdeva-2017} thus a general recommendation for DBM is to use radial distances beyond 20 R$_{\odot}$. Additionally, in many cases constant $w$ and $\gamma$ may also not be valid like in CME-CME interaction events \citep{Temmer-2012, Temmer-2017} or during the passage of high speed solar wind streams \citep{Vrsnak-2010}. 

One of the purposes of DBEM was to take into account the uncertainties regarding $\gamma$ and $w$ that are difficult to measure or estimate directly. In one of the first DBEM versions this was performed by creating a certain number $m$ of synthetic measurements for $\gamma$ and $w$ assuming that the real measurements of these parameters follow a normal distribution and the density of these $m$ measurements is larger around the mean value (for more information see \citealt{Dumbovic-2018}). Each set of such synthetic measurements was then permuted together with the CME input ensembles to obtain an ensemble of DBEM results. 

Based on the methods used by \citet{Dumbovic-2018} to create synthetic measurements, a DBEMv1 (version 1) web online tool was created, which does not use CME ensembles, but instead creates them. The advantage of such tool over the procedure described in \citet{Dumbovic-2018} is that the ensemble input can be created by one observer using a single measurement method. Thus, the DBEMv1 web tool had synthetic measurements generated for all six input parameters (including CME launch time, speed, angular width and longitude) and the number of final DBEM runs was more than 500 000 when the number of synthetic measurements, $m$ was equal or larger than 9 per single input parameter. Such large number of DBEM runs was required in order to produce reliable results without certain biases and thus, more CPU power and time was needed for the calculations of DBEMv1. 

\begin{figure}
   \centering
   \includegraphics[width=0.8\textwidth]{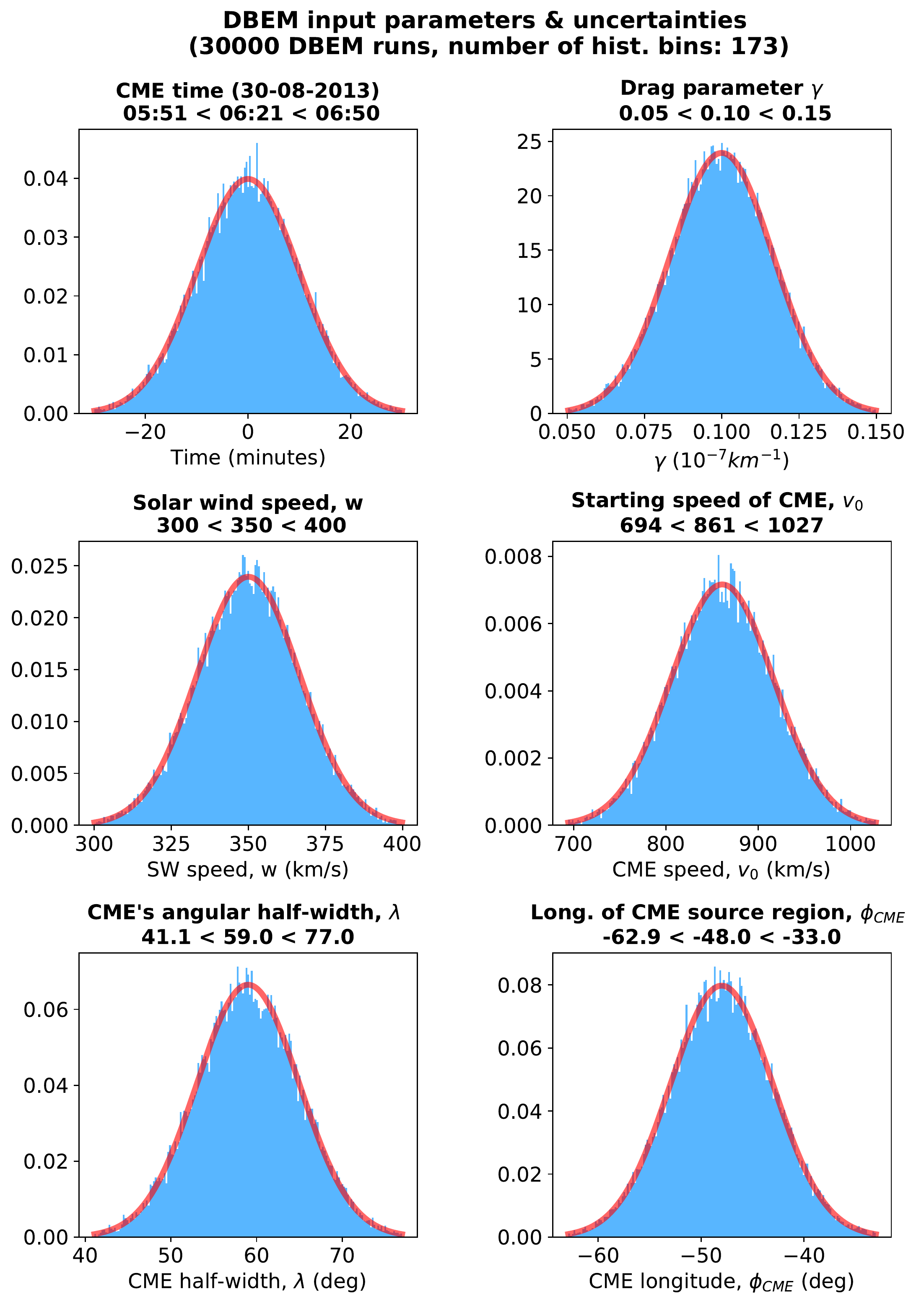}
      \caption{Example of the six input parameters with their corresponding uncertainties for the CME launched on 30 August 2013 at 6:21 UTC shown in histograms as probability density functions (PDFs). Given uncertainties (99.7\% confidence intervals or 3$\sigma$) are indicated above each plot (lower uncertainty $<$ input value $<$ upper uncertainty). The blue bars represent the DBEM samples (in this case 30 000 for each parameter) and the red solid line denotes the calculated normal distribution. Since DBEMv2, this figure is produced automatically for all input parameters in every DBEM calculation. }
         \label{Fig1-DBEM-input}
   \end{figure}

To resolve the mentioned issues, DBEMv2 (version 2) was developed where an ensemble is produced using random values that follow a normal distribution. This can be done under the assumption that the real measurement of input parameters $X_i$ follow a normal distribution and $X_i=\bar{x}_i\pm\Delta x_i$, where $\bar{x}_i$ is the mean of the normal distribution and the uncertainty is equal to three standard deviations. In this case $\Delta x_i=3\sigma$ defines a range around the mean value where $99.7\%$ of measurements are included. For each input parameter with uncertainty (CME launch time $t_0$, speed $v_0$, angular width $\lambda$  and longitude $\phi_{\mathrm{CME}}$, drag parameter $\gamma$ and solar wind speed $w$), the random samples are drawn from a normal distribution determined by $\bar{x}_i$ and $\sigma$ where the number of samples correspond to the number of the ensemble members or DBEM runs. The samples for each input parameter define the probability density functions, PDFs and the example of PDFs for one CME and all input parameters are shown in Figure \ref{Fig1-DBEM-input}. To be able to run the DBEM, user has to define all needed input parameters, $\bar{x}_i$ ($t_0$, $v_0$, $\lambda$, $\phi_{\mathrm{CME}}$, $\gamma$, $w$) and their corresponding uncertainties given as $\pm\Delta x_i$.

In this way the DBEMv2 input distributions or PDFs are better represented than in DBEMv1. Furthermore, the DBEMv2 with the random samples converges to stable results much faster than DBEMv1 with synthetic measurements, which allows to have lower number of DBEM runs and increased the DBEM calculation speed. Additionally, DBEMv2 web tool version has clear advantage over DBEMv1 that the user can chose the exact number of DBEM runs, where the number of runs in the DBEMv1 web tool was determined by a number of synthetic measurements ($m^n$, where $m$ is number of synthetic measurement and $n$ is the number of input parameters). When compared to observations, DBEMv2 also performed slightly better than DBEMv1 and more detailed comparison is given in Section \ref{DBEM-comparison}.

\begin{figure}
   \centering
   \includegraphics[width=0.6\textwidth]{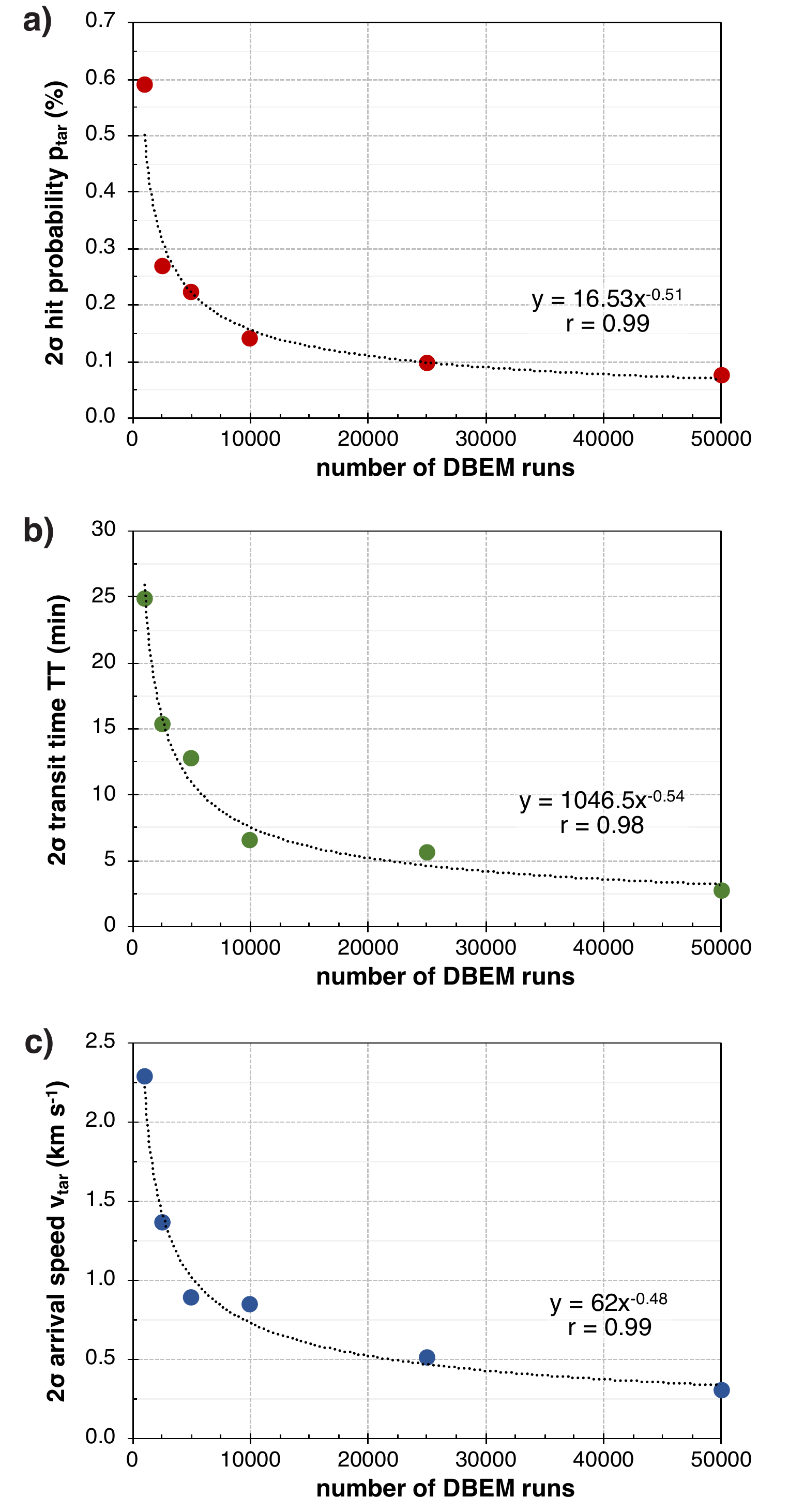}
      \caption{Dependence of standard deviations in DBEM results due to randomly generated input probability density functions on the number of performed DBEM runs calculated for: (a) probability that CME hits the target, $p_{\mathrm{tar}}$ (\%), (b) CME transit time, TT (minutes) and (c) CME arrival speed at target, $v_{\mathrm{tar}}$ (km\,s$^{-1}$). Red ($p_{\mathrm{tar}}$), green (TT) and blue ($v_{\mathrm{tar}}$) points represent standard deviations given as 2$\sigma$ (95\% confidence interval) and calculated by running 20 times DBEM with the identical input data but the different number of runs. The best power law fit is denoted with the black dotted curve.  }
         \label{Fig2-convergence}
   \end{figure}

Due to random input, DBEMv2 produces every time slightly different results, but the differences converge with increasing number of DBEM runs. To test the convergence of the DBEMv2 results, the model was run for 20 times using the same input parameters allowing to calculate the standard deviation for different number of DBEM runs and each obtained output parameter. The convergence results depending on the number of DBEM runs are shown in Figure \ref{Fig2-convergence} for the probability that a CME hits the target $p_{\mathrm{tar}}$ (detailed description is given at the end of this Section), transit time TT and arrival speed $v_{\mathrm{tar}}$. All three output parameters deviations follow a power law with increasing number of DBEM runs (denoted by black dotted lines) and already for 10,000 DBEM runs the results almost fully converge. The deviations are expressed as 2$\sigma$ (95\% confidence interval) and for 10,000 DBEM runs and hit probability, $p_{\mathrm{tar}}$ are just 0.14\% (Figure \ref{Fig2-convergence}a). The same deviations for transit time (TT) are 6 minutes (Figure \ref{Fig2-convergence}b) and for arrival speed, $v_{\mathrm{tar}}$ are 0.85 km\,s$^{-1}$ (Figure \ref{Fig2-convergence}c). This is several orders smaller and thus negligible compared to the other known model uncertainties and errors. 

DBEMv2 was updated with DBEMv25 (version 2.5) where several improvements were implemented. The new routine for transit time calculation was developed allowing also to take into account the proper target motions (\eg planets and satellites including Earth) during the CME passage from the Sun to the target. The DBEM ephemerides data were changed to JPL HORIZONS system\footnote{See \url{https://ssd.jpl.nasa.gov/horizons.cgi}.} that enabled to add the new targets more easily (\eg Solar Orbiter, Parker Solar Probe etc.). To increase the DBEM calculation speed, the DBEM Python code was further parallelized to allow calculation on multiple CPU (\eg on the current server and 32 CPU cores it is possible to run 100,000 DBEM runs in few seconds). However, the main routines for DBEM calculation and its visualisation were left unchanged from DBEMv2 to DBEMv25 except for the correction of several minor bugs. 

\begin{figure}
   \centering
   \includegraphics[width=\textwidth]{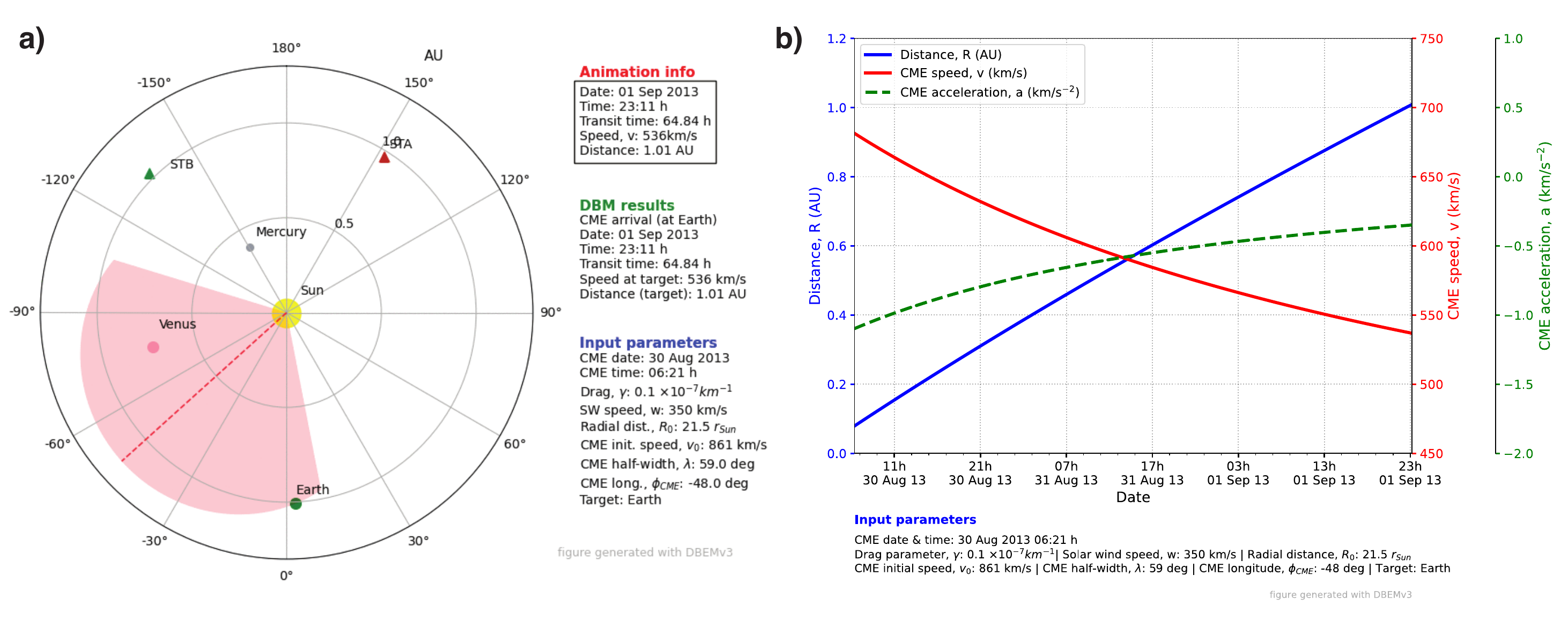}
      \caption{Example of DBM visualization implemented in DBEMv3 for CME launched on 30 August 2013 at 6:21 UTC. a) CME geometry plot where the CME is marked with red shaded area and CME apex with a red dashed line. b) CME kinematic plot with CME distance, $R$ (blue solid line), CME speed, $v$ (red solid line) and CME acceleration, $a$ (green dashed line) plotted against the time (in UTC) where the first point on x-axis corresponds to CME launch time and the last point to time when CME hits the target (Earth).}
         \label{Fig3-DBM}
\end{figure}

Since DBEM already uses the DBM module to calculate the results, it was practical to integrate the current DBM web tool (available at the Space Situational Awareness (SSA) European Space Agency (ESA) portal\footnote{See \url{http://swe.ssa.esa.int/heliospheric-weather}.}) into a new version of DBEMv3 (version 3). In such way, after providing input parameters and before giving all required uncertainties, the user can get the basic DBM results and CME propagation visualizations. One example for such visualization is given in Figure \ref{Fig3-DBM} showing the CME geometry (Figure \ref{Fig3-DBM}a) together with all information regarding the DBM results and input parameters as well as CME kinematics (Figure \ref{Fig3-DBM}b) for CME distance $R$, speed $v$ and acceleration $a$ during the transit time. Thus, the main changes in DBEMv3 compared to DBEMv25 are in the terms of improved visualizations and integration of DBM tool as well as the implementation of the new Graduated Cylindrical Shell (GCS) model option \citep{Thernisien-2006, Thernisien-2009, Thernisien-2011}. GCS is an empirical model to represent the flux rope structure of CMEs and it was recently chosen as the recommended model for CME measurements \citep{Verbeke-2019}. It assumes that geometrically a CME can be described as a hollow croissant consisting of a tubular section forming the main body of the structure which is attached to two cones that correspond to the CME "legs". GCS provides an improved cross-section derivation for calculating more reliably the CME propagation in the ecliptic plane.

In DBEMv3, the GCS model option allows to calculate CME's angular half-width ($\lambda$) from GCS parameters: angular half-width between the ‘‘legs’’ of the GCS model ($\alpha$), CME aspect ratio ($\kappa$) and tilt angle ($\gamma$) by using following equations and Equation 4 derived by \citet{Dumbovic-2019}:

\begin{eqnarray}
      \kappa &=& \mathrm{sin}(\delta) \\
	\omega_{\mathrm{FO}} &=& 2(\alpha + \delta) \\
	\omega_{\mathrm{EO}} &=& 2\delta \\
	\lambda &=& \omega_{\mathrm{FO}} - (\omega_{\mathrm{FO}} - \omega_{\mathrm{EO}})( |\gamma| /90)
\end{eqnarray}

where $\omega_{\mathrm{FO}}$ and $\omega_{\mathrm{EO}}$ are face-on and edge-on widths according to \citet{Thernisien-2011}, $\gamma$ is the tilt angle of croissant axis.

\begin{figure}
   \centering
   \includegraphics[width=\textwidth]{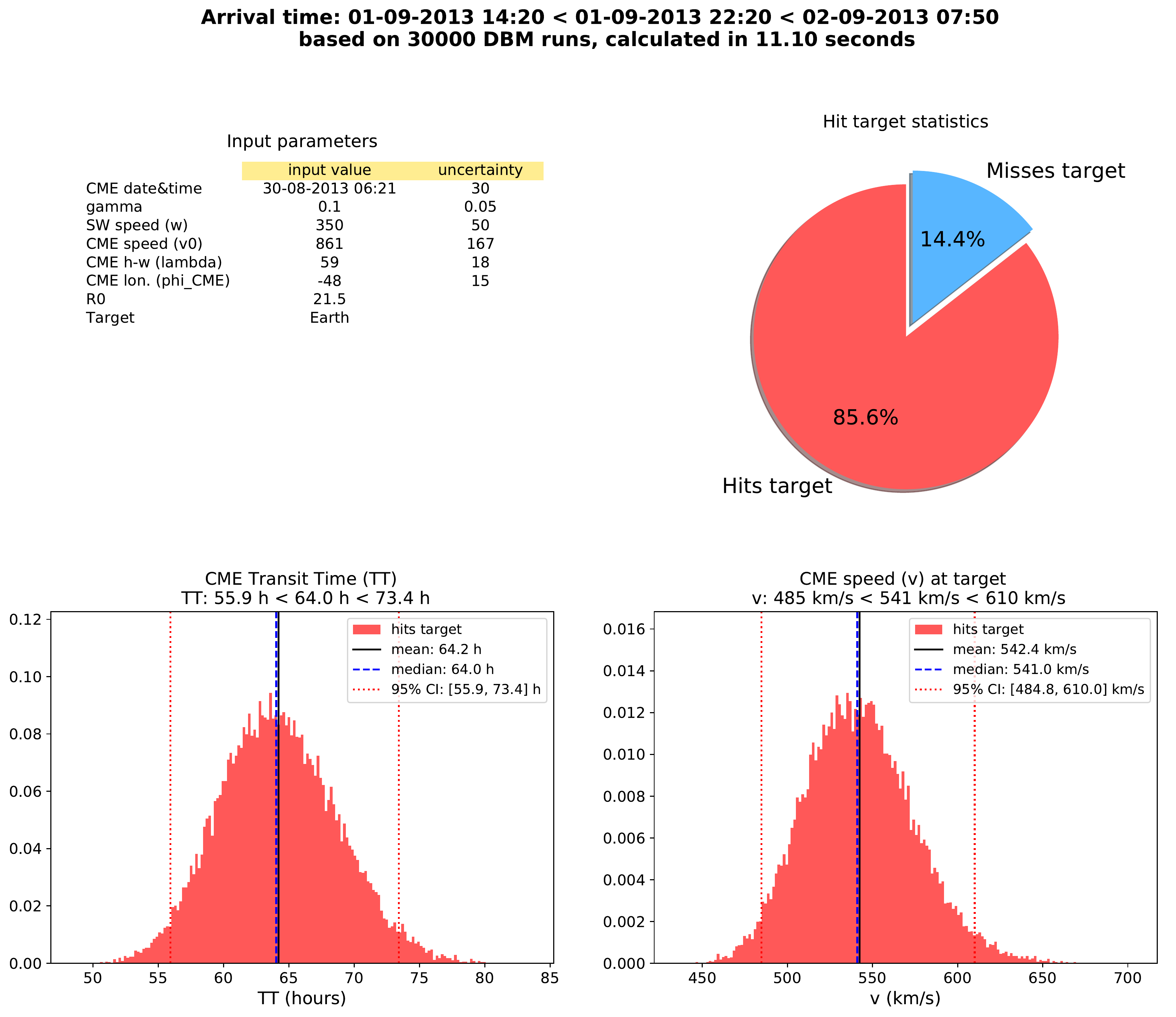}
      \caption{Example of DBEM results for CME launched on 30 August 2013 at 6:21 UTC. The upper left panel shows table with all input parameters and their uncertainties. The upper right panel shows pie chart with the hit target statistics (miss or hit). Lower panels show histograms with transit time, TT (left) and CME speed at target, $v_{\mathrm{tar}}$ (right). Median value is denoted in histograms as black solid line and mean value as blue dashed line. The calculated 95\% confidence intervals are shown by the red dotted lines.}
         \label{Fig4-DBEM-results}
\end{figure}

After providing the required uncertainties for all six input parameters (CME launch time, $\gamma$, $w$, $v_0$, $\lambda$, $\phi_{\mathrm{CME}}$) DBEMv3 calculates probabilistic arrival time, transit time TT, probability that CME hits target $p_{\mathrm{tar}}$, CME speed $v_{\mathrm{tar}}$ and acceleration $a_{\mathrm{tar}}$ at a target. The example of DBEMv3 output is given in Figure \ref{Fig4-DBEM-results} and shows the same event as in Figures \ref{Fig1-DBEM-input} and \ref{Fig3-DBM}. The upper left panel in Figure \ref{Fig4-DBEM-results} shows all input parameters with their uncertainties for a given DBEM calculation. The right upper panel in Figure \ref{Fig4-DBEM-results} provides a pie chart for probability of CME arrival ($p_{\mathrm{tar}}$). DBEM calculates for each ensemble member whether the CME will hit or miss the target (in this case Earth). For the whole ensemble the probability of CME arrival at target $p_{\mathrm{tar}}$ is calculated as ratio between the number of ensemble members that hit the target, $n_{\mathrm{hit}}$ and total number of all ensemble members, $n_{\mathrm{tot}}$ or $p_{\mathrm{tar}} = n_{\mathrm{hits}}/n_{\mathrm{tot}}$. The lower left panel in \ref{Fig4-DBEM-results} shows the distribution for transit time, TT based on the DBEM ensemble members that are calculated to hit the target (\ie red part in the pie chart in the upper right panel of Figure \ref{Fig4-DBEM-results}). From this distribution it is possible to determine main distribution parameters as mean and median values and 95th percentiles as 95\% confidence intervals (CI). Analogue to TT, the distribution for CME speed at target, $v_{\mathrm{tar}}$ is given in the lower right panel of Figure \ref{Fig4-DBEM-results} with the same statistics parameters (mean, median, 95\% CI). In this way DBEM provides for all output parameters (TT, $v_{\mathrm{tar}}$, $a_{\mathrm{tar}}$) the expected range given by the 95\%  CIs and median as the most likely value, what is also denoted above each histogram (lower 95\% CI $<$ median $<$ upper 95\% CI). If DBEM calculates that all ensemble members will miss the target (0\% chance to hit the target), it won't be able to provide CME arrival time and speed as well as the corresponding histogram plots with statistics.  

\section{DBEMv3 and DBEMv1 Comparison}  \label{DBEM-comparison}

To compare DBEMv3 with previous DBEMv1 model version, the identical CME sample as in \citet{Dumbovic-2018} was employed, reduced to altogether 16 CME-ICMEs pairs (or events that hit the Earth) and denoted as sample 1 (see Table \ref{table1-CME-list}). Here should be noted that the DBEM employed in \citet{Dumbovic-2018} using the ensembles of CME measurements as input is slightly different in terms of input from the DBEMv1 web tool which creates the CME ensembles by itself by using the uncertainties for CME properties parameters as input. In \citet{Dumbovic-2018} and \citet{Mays-2015} the ensemble of CME measurements was prepared using the STEREOCAT tool by employing the multiple measurements from coronagraph images. Thus, for DBEMv1 evaluation purposes presented here, the input CME  parameters (CME launch speed, half-width and longitude) and their corresponding uncertainties were determined as median values from the ensemble of the CME measurements in \citet{Dumbovic-2018}. 

\begin{table}
\centering
{\begin{minipage}{\linewidth}List of 25 CME-ICMEs used for DBEMv3 evaluation and comparison with DBEMv1. First eight columns are DBEM input parameters: CME date and time, CME launch speed ($v_0$) with uncertainty ($\pm\Delta v_0$), CME's angular half-width ($\lambda$) with uncertainty ($\pm\Delta\lambda$) and longitude of CME source region ($\phi_{\mathrm{CME}}$) with uncertainty ($\pm\Delta \phi_{CME}$). For all events, CME start time was set to $\pm\Delta t_0 = 30$ min, the drag parameter ($\gamma$) was set to $0.1 \times 10^{-7}$ km$^{-1}$ with uncertainty, $\pm\Delta \gamma$ = 0.05 $\times 10^{-7}$ km$^{-1}$, solar wind speed ($w$) to 350 km\,s$^{-1}$ with uncertainty, $\pm\Delta w$ = 50 km\,s$^{-1}$, CME starting radial distance, $R_0$ to 21.5 R$_{\odot}$ and target to Earth. The DBEM results are represented with next three columns: probability of arrival ($p_{\mathrm{tar}}$), DBEMv3 calculated transit time (TT$_{\mathrm{DBEM}}$) and CME arrival speed ($v_{\mathrm{tar}}$).  In-situ measurements of CME arrival are shown in the next two columns: observed transit time (TT$_{\mathrm{OBS}}$) and transit tme prediction error ($\Delta$TT$_{\mathrm{err}} =$ TT$_{\mathrm{OBS}} - $TT$_{\mathrm{DBEM}}$). Last column denotes the different samples used for the validation and only the sample 1 corresponds to CME-ICME sample used in \citet{Dumbovic-2018}. \hspace{\textwidth} \end{minipage}}
\caption{}
\label{table1-CME-list}
\resizebox{\textwidth}{!}{%
\begin{tabular}{l c c c c c c c | c c c | c c | c} 
\multicolumn{8}{c}{\textbf{DBEM input}} & \multicolumn{3}{c}{\textbf{DBEM results}} & \multicolumn{2}{c}{\textbf{Observed}} & \\
\hline
\multicolumn{2}{c}{CME start} & $v_0$ & $\pm\Delta v_0$ & $\lambda$ &  $\pm\Delta\lambda$ &  $\phi_{\mathrm{CME}}$ & $\pm\Delta \phi_{CME}$ & $p_{\mathrm{tar}}$ & TT$_{\mathrm{DBEM}}$ & $v_{\mathrm{tar}}$ & TT$_{\mathrm{OBS}}$ & $\Delta$ TT$_{\mathrm{err}}$ & sample\\
date & time & (km\,s$^{-1}$) & (km\,s$^{-1}$) & (deg) & (deg) & (deg) & (deg) & (\%) & (h) & (km\,s$^{-1}$) & (h) & (h) & \\
\hline
03/04/2010&15:00&850&85&41&4&-4&5&100&53.14&604&41.43&11.71&2\\
01/08/2010&12:00&1200&120&53&5&-29&5&100&44.78&684&53.68&-8.90&2\\
15/02/2011&06:00&1000&100&36&4&-11&5&100&48.25&647&67.50&-19.25&2\\
03/08/2011&17:00&1100&110&42&4&23&5&100&46.93&664&48.85&-1.92&2\\
14/09/2011&06:00&500&50&40&4&19&5&100&82.51&442&69.72&12.79&2\\
26/10/2011&18:00&480&48&38&4&34&5&100&96.66&388&88.02&8.64&2\\
26/11/2011&11:00&1000&100&63&6&56&5&100&55.99&587&58.83&-2.84&2\\
15/03/2013&10:00&1100&110&52&5&9&5&100&44.27&691&43.98&0.29&2\\
11/04/2013&10:37&1000&337&55&12&-15&14&100&48.40&645&62.82&-14.42&1\\
21/06/2013&04:51&2006&192&59&11&-48&14&94.55&35.27&835&48.65&-13.38&1\\
30/08/2013&06:21&861&167&59&18&-48&15&85.57&64.11&540&71.13&-7.02&1\\
30/09/2013&01:45&1000&464&66&22&30&20&100&50.02&635&52.58&-2.57&1\\
06/10/2013&18:15&748&268&16&8&2&5&100&59.77&563&53.02&6.76&1\\
07/01/2014&19:48&2399&661&64&13&38&28&99.71&27.89&991&49.25&-21.36&1\\
30/01/2014&20:04&843&388&45&23&-29&40&80.11&58.91&562&78.93&-20.03&1\\
12/02/2014&10:09&740&265&59&17&6&6&100&58.01&564&79.12&-21.11&1\\
18/02/2014&05:20&850&272&52&27&-44&43&62.06&60.99&551&49.28&11.70&1\\
19/02/2014&19:29&884&341&29&14&-10&4&100&53.20&603&86.15&-32.95&1\\
25/02/2014&03:14&1500&776&80&9&-78&25&51.18&38.53&764&62.68&-24.16&1\\
23/03/2014&08:42&716&195&55&17&-60&30&23.28&74.17&483&63.57&10.60&1\\
02/04/2014&15:56&1528&462&51&21&-55&5&20.28&43.09&710&68.40&-25.31&1\\
18/04/2014&15:21&1384&262&46&15&9&4&100&38.05&775&45.18&-7.13&1\\
04/06/2014&22:59&580&162&50&25&-28&37&88.88&78.19&461&72.40&5.79&1\\
19/06/2014&23:57&570&209&44&19&-20&13&99.68&78.75&463&73.27&5.48&1\\
15/03/2015&06:45&817&82&75&8&30&5&100&55.97&584&46.00&9.97&2\\
\hline\\
\end{tabular}}
\end{table}

To further increase the number of analysed CMEs and extend the period of analysis, an additional sample of 9 CME-ICME was also employed in the DBEMv3 evaluation (the sample 2 in Table \ref{table1-CME-list}). However, several DBEM input uncertainties for the CME sample 2 were set as fixed values and were not determined from CME ensembles. For this reason, these samples were separated. Thus, the sample 2 uncertainties for CME initial speed ($v_0$) and CME half-width ($\lambda$) were set to $\pm10\%$ of initial input parameter value and for CME longitude ($\phi_{\mathrm{CME}}$) to $\pm5$ deg (see Table \ref{table1-CME-list}). The rest of DBEM input parameters and uncertainties was the same for both samples: uncertainty for CME start time was set to $\pm\Delta t_0 = 30$ min, the drag parameter ($\gamma$) value to 0.1 $\times 10^{-7}$ km$^{-1}$ with uncertainty, $\pm\Delta \gamma = 0.05 \times 10^{-7}$ km$^{-1}$, solar wind speed ($w$) value to 350 km\,s$^{-1}$ with uncertainty, $\pm\Delta w = 50$ km\,s$^{-1}$, CME starting radial distance, $R_0$ to 21.5 R$_{\odot}$ and target to Earth. Table \ref{table1-CME-list} shows all 25 analyzed CMEs (sample 1 and 2) together with CME initial input parameters and uncertainties ($t_0$, $v_0$, $\lambda$, $\phi_{\mathrm{CME}}$), DBEM results ($p_{\mathrm{tar}}$, TT$_{\mathrm{DBEM}}$, $v_{\mathrm{tar}}$) and in-situ measurements (TT$_{\mathrm{OBS}}$, $\Delta$TT$_{\mathrm{err}}$). 

\begin{figure}
   \centering
   \includegraphics[width=0.75\textwidth]{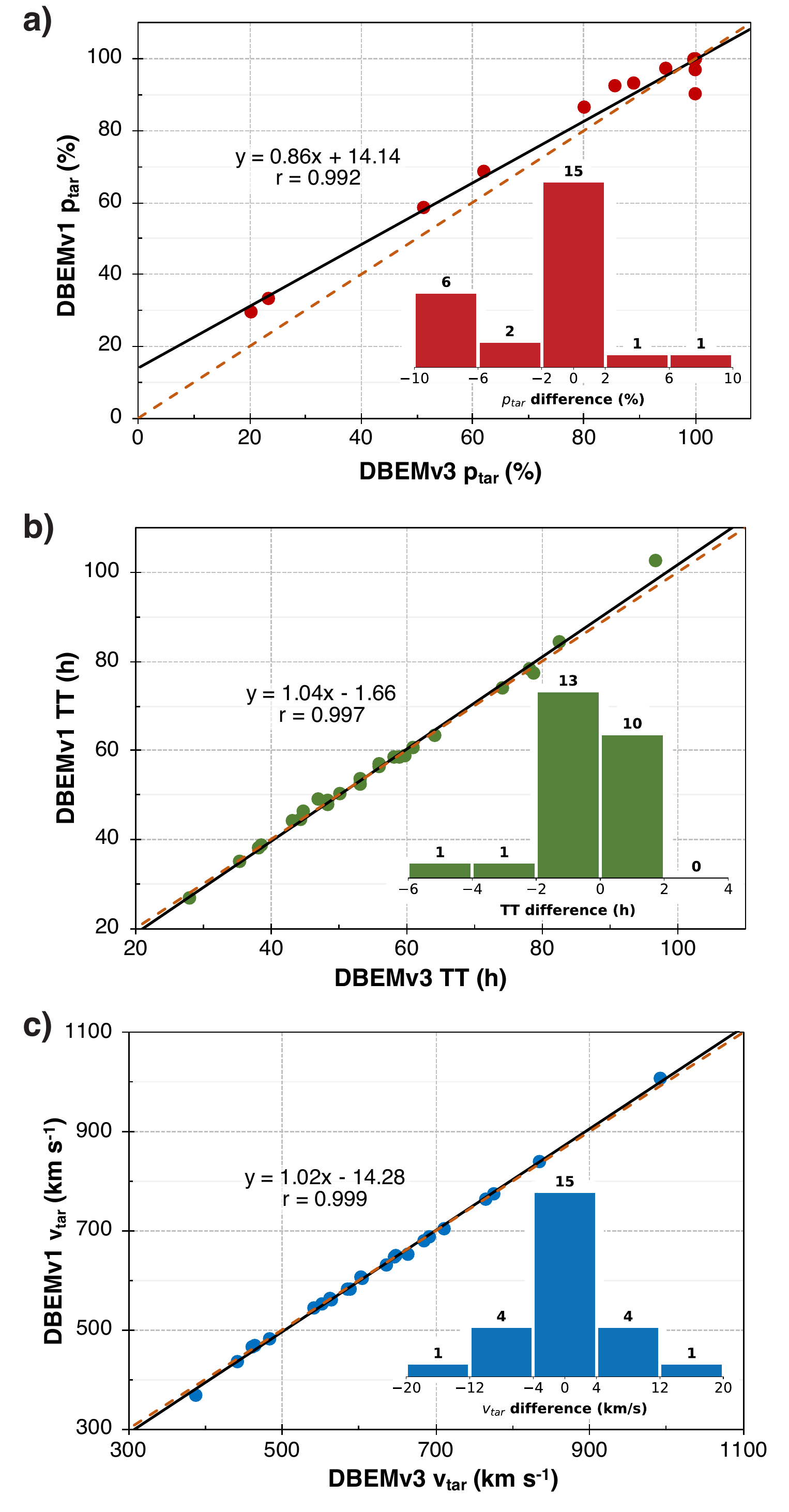}
      \caption{Comparison of DBEMv3 and DBEMv1 using 25 CMEs (sample 1 and 2) for a) probability of CME arrival at the target, $p_{\mathrm{tar}}$ expressed in \%, b) CME transit time, TT expressed in hours and c) CME arrival speed at target, $v_{\mathrm{tar}}$ expressed in km\,s$^{-1}$. The black solid lines show the best linear fit together with the linear fit equation and correlation coefficient, $r$. The orange dashed line represents the identity line (perfect match). Histograms in the insets show the relative differences between DBEMv3 and DBEMv1 where DBEMv1 values are subtracted from DBEMv3 values. The count of events in each bin is shown by the number on top of bins.  }
         \label{Fig5-validation}
   \end{figure}

A comparison of the DBEMv3 and DBEMv1 results are shown in Figure \ref{Fig5-validation}. The probability of CME arrival at target, $p_{\mathrm{tar}}$ in \% is plotted in Figure \ref{Fig5-validation}a as scatter where calculated DBEMv1 $p_{\mathrm{tar}}$ is on \textit{y-axis} and DBEMv3 $p_{\mathrm{tar}}$ on \textit{x-axis}. The provided linear fit (black solid line) is very close to the identity line or perfect match (orange dotted line) with the correlation coefficient, $r$ of 0.992. The histogram on the right side of the plot shows the relative differences between DBEMv3 and DBEMv1 where DBEMv1 $p_{\mathrm{tar}}$ is subtracted from DBEMv3 $p_{\mathrm{tar}}$ value. About 60\% of the analyzed CMEs have very small difference and $p_{\mathrm{tar}}$ is within $\pm 2\%$ and 25\% CMEs have somewhat larger but still rather small difference for $p_{\mathrm{tar}}$ between 6\% and 10\%. The smaller $p_{\mathrm{tar}}$ values obtained for DBEMv3 are mainly due to the implementation of new DBM routine since the DBEMv25 which takes into account the target movements during the CME transit. The proper target movement is more important when CME hits the target near the flank and resulting in some cases with the miss due to the target moment. Finally during many DBM runs this reflects in the somewhat lower calculated $p_{\mathrm{tar}}$ values what can be observed from Figure \ref{Fig5-validation}a that the larger differences in $p_{\mathrm{tar}}$ are more pronounced for smaller $p_{\mathrm{tar}}$ values. A very small differences were obtained also for the CME transit time, TT expressed in hours as shown in Figure \ref{Fig5-validation}b plotted in the same way as for $p_{\mathrm{tar}}$. In this case linear fit for TT is better than for $p_{\mathrm{tar}}$ and it is almost on the perfect match line with the correlation coefficient, $r$ of 0.997. As shown in the TT difference histogram, nearly all or 92\% of all differences in TT are within $\pm 2$ h which is very small compared to other model errors and uncertainties. The comparison for the CME arrival speed at target, $v_{\mathrm{tar}}$ is given in Figure \ref{Fig5-validation}c and here the differences are also very small with the highest $r$ of 0.999. The corresponding histogram shows normal distribution of $v_{\mathrm{tar}}$ difference values and 60\% of them are within the value of $\pm 4$ km\,s$^{-1}$.

\begin{figure}
   \centering
   \includegraphics[width=0.70\textwidth]{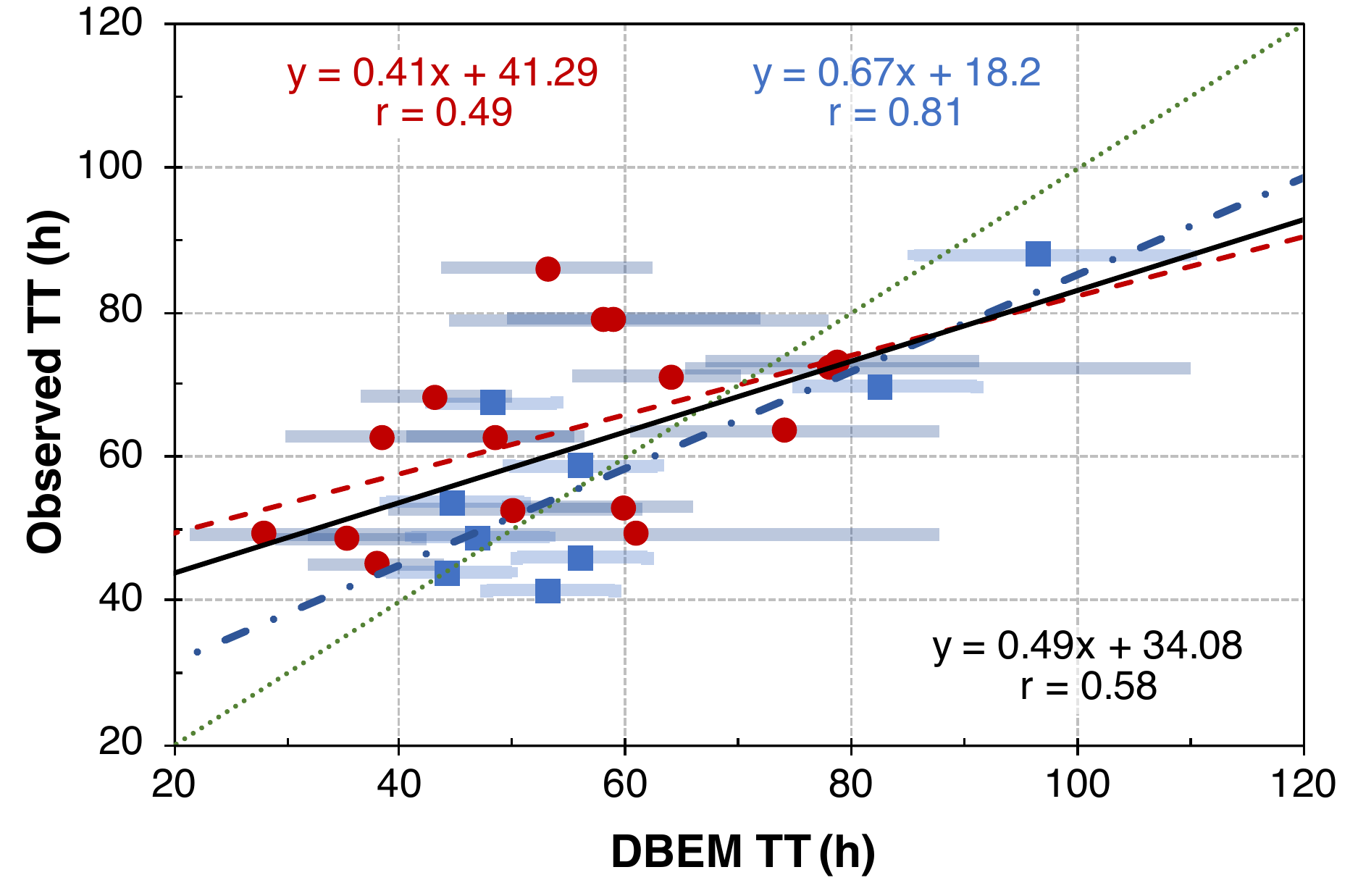}
      \caption{ Observed \textit{vs.} calculated DBEMv3 transit time, TT expressed in hours and DBEM TT values are given as medians of the corresponding distributions. Blue bars around points are DBEMv3 TT 95\% confidence intervals. Red dots represent the CME sample 1 and blue rectangles the sample 2. The corresponding linear fits with their equations and correlation coefficients, $r$ are also shown on the plot where red dashed line is for the sample 1, blue dashed dotted line for the sample 2 and black solid line for both samples together. The green dotted line represents the identity line (perfect match). }
         \label{Fig6-obs-DBEM}
   \end{figure}

\begin{figure}
   \centering
   \includegraphics[width=0.65\textwidth]{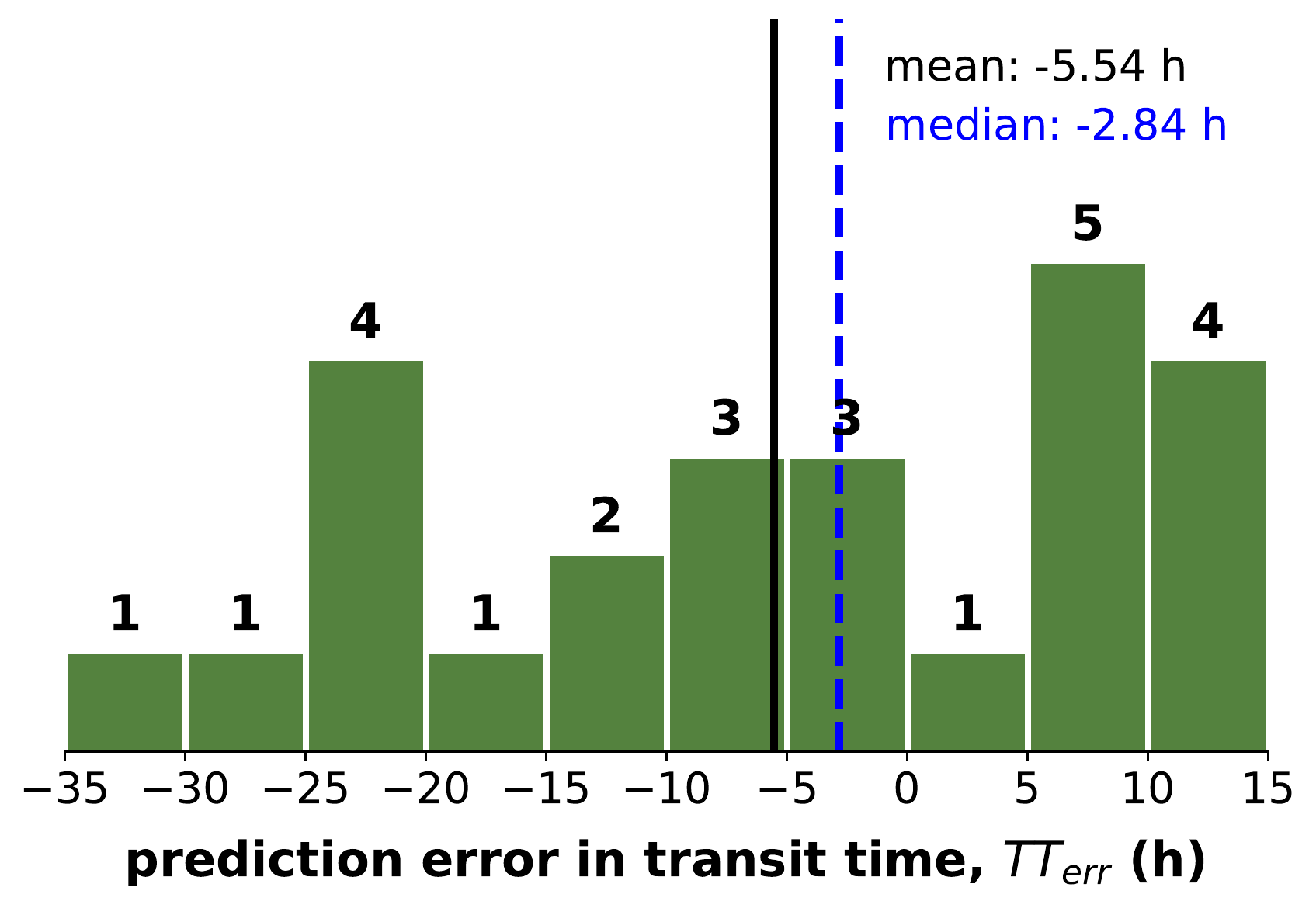}
      \caption{ Transit time prediction error, $\Delta$TT$_{\mathrm{err}}$ expressed in hours for DBEMv3 and 25 CMEs (sample 1+2). The mean and median values are represented by blue dashed and black solid line. On the top of bins the count of events in each bin is indicated. }
         \label{Fig7-TT-obs-hist}
   \end{figure}

The DBEMv3 evaluation is shown in Figure \ref{Fig6-obs-DBEM} as scatter plot between observed TT (\textit{y-axis}) and predicted DBEMv3 TT (\textit{x-axis}). The sample 1 is represented with red dots and the blue bars are DBEM TT 95\% confidence intervals calculated from TT distributions (an example is shown in Figure \ref{Fig4-DBEM-results}). The corresponding linear fit for the sample 1 is denoted with red dashed line with correlation coefficient  $r=0.49$. In comparison, the sample 2 (blue rectangles) shows much better agreement with the observed TT and identity line (green dotted line) where its linear fit (blue dashed dotted line) has much higher $r=0.81$. As expected, the linear fit for both samples 1 and 2 (solid black line) is somewhere in between, with $r=0.54$. The DBEMv3 transit time prediction error, $\Delta$TT$_{\mathrm{err}}$ in hours is shown in Figure \ref{Fig7-TT-obs-hist} as histogram for all 25 CMEs together (sample 1+2). $\Delta$TT$_{\mathrm{err}}$ is calculated as difference between DBEM transit time, TT$_{\mathrm{DBEM}}$ and in-situ observation of transit time, TT$_{\mathrm{OBS}}$ ($\Delta$TT$_{\mathrm{err}}=$ TT$_{\mathrm{DBEM}} -$TT$_{\mathrm{OBS}}$). From the histogram shown in Figure \ref{Fig7-TT-obs-hist}, it is clear that the majority of the $\Delta$TT$_{\mathrm{err}}$ values are shifted towards negative values resulting also in a negative mean error (ME) of $-5.54$ h and median value of $-2.84$ h. Despite that fact,  DBEMv3 still shows reasonable performance compared to other models \citep{Dumbovic-2018, Riley-2018}, since 18 out of 25 CMEs (72\%) are still within a reasonable range of $\pm15$ h, in about one third of the cases under study, DBEM seems to largely underestimate the TT and from Figure \ref{Fig7-TT-obs-hist} it can be seen that 6 CMEs have $\Delta$TT$_{\mathrm{err}}$ larger than $-15$ h with maximal value for one CME of almost $-33$ h.

In Table \ref{table2-DBEM-stats} different DBEMv3 and DBEMv1 model errors calculated from $\Delta$TT$_{\mathrm{err}}$ for the sample 1 and 2 are compared. For comparison purposes the values written in bold in Table \ref{table2-DBEM-stats} mark the better performance of DBEMv3 compared to the DBEMv1 (values in brackets) and majority of errors show indeed a slightly better performance of the improved DBEMv3 version. However, in some cases DBEMv1 provided smaller TT errors than DBEMv3, but these differences are small and in most cases smaller than half hour error in TT. Nevertheless, it should be noted that in the both cases, when the DBEMv3 was performing better than DBEMv1 or worse, the order of magnitude of the change in model errors ($\Delta$TT$_{\mathrm{err}}$) related to different DBEM versions is much smaller than other uncertainties in measurements and known limitations in the DBM model itself.

\begin{table}
\caption{  Comparison of various model errors in hours calculated from transit time prediction error, $\Delta$TT$_{\mathrm{err}}$ for 25 CMEs (sample 1+2). The first value in each cell is given for DBEMv3 and the second value in brackets for DBEMv1. The better performance of DBEMv3 compared to DBEMv1 is denoted with bold values. ME is the mean error, MAE is the mean absolute error, RMSE is the root mean square error and $\sigma$ is the calculated standard deviation. Note: slightly different calculated values for DBEMv1 ME, MAE and RMSE compared to the ones published in \citet{Dumbovic-2018} are due to the rounding errors and slightly different number of DBEMv1 runs that were performed in this study.}
\label{table2-DBEM-stats}                     
\begin{tabular}{l | c c c}        
\hline                 
error & sample 1 (h) & sample 2 (h) & sample 1+2 (h) \\    
\hline                        
ME&\textbf{-9.32} (-9.59)&\textbf{1.17} (2.52)&-5.54 (-5.23)\\ 
MAE&14.36 (14.26)&\textbf{8.48} (9.02)&\textbf{12.24} (12.37)\\ 
RMSE&\textbf{16.71} (16.75)&\textbf{10.2} (11.16)&\textbf{14.7} (14.98)\\ 
median &\textbf{-10.26} (-10.72)&\textbf{0.29} (0.44)&-2.84 (-2.23)\\ 
min&\textbf{-32.95} (-33.66)&\textbf{-19.25} (-19.67)&\textbf{-32.95} (-33.66)\\ 
max&11.7 (11.44)&\textbf{12.79} (14.51)&\textbf{12.79} (14.51)\\ 
$\sigma$&14.32 (14.19)&\textbf{10.75} (11.53)&\textbf{13.89} (14.33)\\ 
\hline                              
\end{tabular}             
\end{table}

The prediction errors for CME arrival speed at target, $v_{\mathrm{tar}}$ were also evaluated with DBEMv3 using the sample 1 and observed in-situ CME speeds listed in Table 1 in \citet{Dumbovic-2018}. The $v_{\mathrm{tar}}$ prediction errors ($\Delta v_{\mathrm{err}}$) were calculated by subtracting the observed CME speed from DBEM prediction ($\Delta v_{\mathrm{err}} = v_{\mathrm{DBEM}} - v_{\mathrm{OBS}}$). With DBEMv3 the $\Delta v_{\mathrm{err}}$ calculated mean error (ME) is 79.7 km\,s$^{-1}$, mean absolute error (MAE) 131.2 km\,s$^{-1}$ and root mean square error (RMSE) 175.6 km\,s$^{-1}$. The DBEMv1 showed similar prediction errors for $v_{\mathrm{tar}}$ (ME: 82 km\,s$^{-1}$, MAE: 132.7 km\,s$^{-1}$, RMSE: 178.6 km\,s$^{-1}$) and compared to DBEMv3 all these errors were slightly reduced (ME: -2.3 km\,s$^{-1}$, MAE: -1.5 km\,s$^{-1}$, RMSE: -3 km\,s$^{-1}$). Single events showed in some cases larger improvements (up to 18 km\,s$^{-1}$) for $\Delta v_{\mathrm{err}}$, but for half of events the errors were also smaller in the case of DBEMv1. Thus, a similar conclusion could be made as in the case of TT predictions errors, that the changes in errors related to DBEM model improvements are rather quite small compared to CME input uncertainties and DBM model errors.

\section{Quantifying the DBEMv3 Errors}  \label{DBEM-errors}

Beside the DBEM version comparison and evaluation, the additional important results can be obtained from Figure \ref{Fig6-obs-DBEM} and Table \ref{table2-DBEM-stats}. It is easy to notice that two samples used for model evaluation can give very different results and thus also the found model errors. In Figure \ref{Fig6-obs-DBEM} a large difference in DBEMv3 performance can be observed for two analyzed samples and the TT prediction for the sample 2  ($r=0.81$) is substantially better than for the sample 1 ($r=0.49$). This is also supported by errors listed in Table \ref{table2-DBEM-stats} where for the sample 1 mean error (ME) is -9.32 h, mean absolute error (MAE) 14.36 h and root mean square error (RMSE) is 16.71 h. For the sample 2 these errors are much lower: ME = 1.17 h, MAE = 8.48 h, and RMSE = 10.2 h. 

One of the explanations for this difference in DBEM prediction errors could be related to the size of the samples where sample 1 is almost twice as large as sample 2 (16 CMEs compared to 9 events). Another probably more important explanation for these differences is how well the selected sample is optimized or is it suitable for model evaluation or not as well as how well CMEs were observed, and how complex the prevailing solar wind structure was. In sample 1, altogether 5 CMEs are identified to have $\Delta$TT$_{\mathrm{err}}$ larger than 20 h and all of these events are during the first half of 2014 when solar cycle 24 was on its maximum. If these 5 CMEs with $\Delta$TT$_{\mathrm{err}}$ larger than 20 h are excluded, the DBEMv3 performance and model errors become very similar to the sample 2 prediction errors, $\Delta$TT$_{\mathrm{err}}$. 

It should be also mentioned that uncertainties and errors in initial CME parameters like CME initial speed can also result in larger model errors. In the case of the CME sample 1, three out of five CMEs with $\Delta$TT$_{\mathrm{err}}$ larger than 20 h are very fast CMEs ($v_0 > 1500$ km\,s$^{-1}$) and it was already argued by \citet{Dumbovic-2018} that CMEs with underestimated TT may be related to overestimation of the CME initial speed for fast CMEs.

On the other hand, sample 2 doesn't contain any CME with $\Delta$TT$_{\mathrm{err}}$ larger than 20 h. Since DBM relies on the assumption that $w$ and $\gamma$ are constant, such assumption is less valid during solar maximum than it is the case during solar minimum \citep{Vrsnak-2014} due to more frequent CME-CME interaction events \citep{Temmer-2012, Temmer-2017, Rodriguez-2020} or appearance of high speed streams \citep{Vrsnak-2010}. The majority of CMEs in sample 2 is from 2011 \ie, belonging to a different solar activity phase than in the case of sample 2 and 2014. For example, in 2014 more than 300 CMEs per month were recorded, whereas in 2011 it was less than 200 CMEs per month \citep{Lamy-2019}. Potentially, this can cause larger prediction errors, $\Delta$TT$_{\mathrm{err}}$ for certain CMEs which could be also the case for sample 1. 

For specific and more complex CMEs proper estimation of $w$ and $\gamma$ can be essential to reduce the model errors and real values can differ very much from values used in the evaluation ($w = 350$ km\,s$^{-1}$, $\pm\Delta w = 50$ km\,s$^{-1}$, $\gamma = 0.1 \times 10^{-7}$ km$^{-1}$, $\pm\Delta \gamma$ = 0.05 $\times 10^{-7}$ km$^{-1}$). To investigate this, in Section \ref{DBEM-gamma-w} the reverse modelling with DBEMv3 is presented to determine the best $w$ and $\gamma$ parameters to obtain $\Delta$TT$_{\mathrm{err}} = 0$.

Considering these issues, it still remains question, what is the best practice to prepare sample for model evaluation \citep{Verbeke-2019}. Obviously, the small CME samples for model evaluation are not the best solution to be compared with other models and various samples used to evaluate various models can result in a vast range in performance characteristics of various models, where \eg MAE can range from 3 to 18 h \citep{Vourlidas-2019}. Model optimized CME samples with less than 10 CMEs in studies like that by \citet{Hess-Zhang-2015} or \citet{Corona-Romero-2015}, can result in very small model errors \eg MAE less than 5 h that will be different (and worse) for larger general samples where MAE will be larger than 10 h like in \citet{Wold-2018} or \citet{Riley-2018}. With that in mind, we analyze a larger and more comprehensive sample (denoted as sample 3) based on the ICME list from \citep{Richardson-Cane-2010} that contains more than hundred CMEs (presented in Section \ref{RC-list-analysis}). 

\section{Reverse Modelling with DBEM to Find Optimal $\gamma$ and $w$ Parameters} \label{DBEM-gamma-w}

For two DBM input parameters, $\gamma$ and $w$, it is difficult to obtain direct measurements. The drag parameter depends on the characteristics of both CME and solar wind such as CME cross sectional area perpendicular to its propagation direction, the ambient solar wind density and the CME mass \citep{Vrsnak-2010}. For example, $\gamma$ is larger for broader, low-mass CMEs in a high-density (slow) solar wind and vice versa. In complex solar wind environments that are usually more pronounced during solar maximum than solar minimum, both parameters $\gamma$ and $w$ may not be constant as assumed in the analytical DBM which may be a reason for errors in TT prediction \citep{Vrsnak-2013}. For example, this may be the case for a fast CME first propagating through slow solar wind and then entering a fast solar wind stream \citep{Temmer-2011} or for the fast CME encountering another slow CME launched earlier in the same direction where the slow CME might be “pushed” by fast CME \citep{Temmer-2012}.

To investigate, how strongly the estimated $\gamma$ and $w$ values in sample 1 taken from \citet{Dumbovic-2018} may be responsible in some cases for the large prediction errors (\eg $\Delta$TT$_{\mathrm{err}}$) and if it is possible to obtain better optimized values to improve the DBEM forecast, reverse modelling with DBEMv3 was performed where the prediction errors were set to zero (TT$_{\mathrm{DBEM}}$ = TT$_{\mathrm{OBS}}$ and $\Delta$TT$_{\mathrm{err}}$ = 0). The similar analysis related to TT prediction and finding optimal $\gamma$ and $w$ parameters was already presented in \citet{Paouris-2021} that compared the DBEM with the empirical CME propagation model EAM. However, here we present this analysis in much greater detail. The task to find optimal $\gamma$ and $w$ parameters was done in two steps.

In the first step, it was necessary to determine the appropriate $\gamma$ and $w$ range for the analyzed sample. The reason for that procedure was that by performing the reverse modelling it is possible to obtain the mathematical solutions for $\gamma$ and $w$ that may be outside of expected realistic and physical ranges for these parameters \citep{Vrsnak-2013}. For that purpose in-situ data from the Solar Wind Experiment, SWE \citep{Ogilvie-1995} onboard the Wind spacecraft \citep{Szabo-2015} was employed to estimate the solar wind speed, $w$ upstream of the shock/sheath and behind the CME trailing edge by using time-series plots. For each event and measured $w$, using DBM the best fit for $\gamma$ was then searched to obtain the perfect TT and also perfect  $v_{\mathrm{tar}}$. Perfect TT was defined as TT within $\pm1$ h of the actual observed TT$_{\mathrm{OBS}}$ so that absolute $|\Delta$TT$_{\mathrm{err}}| \le 1$ h. In the same way the perfect arrival speed was defined as $v_{\mathrm{tar}}$ within $\pm10$ km\,s$^{-1}$ of the actual observed $v_{\mathrm{tar}}$ (|$\Delta v_{\mathrm{err}}| \le 10$ km\,s$^{-1}$). Therefore, for sample 1 we constrained the parameter space of possible $\gamma$ and $w$ values to  $\gamma = 0.3 \times 10^{-7}$ km$^{-1}$ with the uncertainty, $\pm\Delta \gamma = 0.29 \times 10^{-7}$ km$^{-1}$ and $w = 450$ km\,s$^{-1}$ with the uncertainty $\pm \Delta w = 150$ km\,s$^{-1}$. The uncertainty for the drag parameter ($\pm\Delta \gamma$) was set as large as possible, but also avoiding to run the model with the zero or negative $\gamma$ values, with the purpose to investigate the whole possible range of $\gamma$ values in relation to the perfect TT and $v_{\mathrm{tar}}$. 

\begin{figure}
   \centering
   \includegraphics[width=\textwidth]{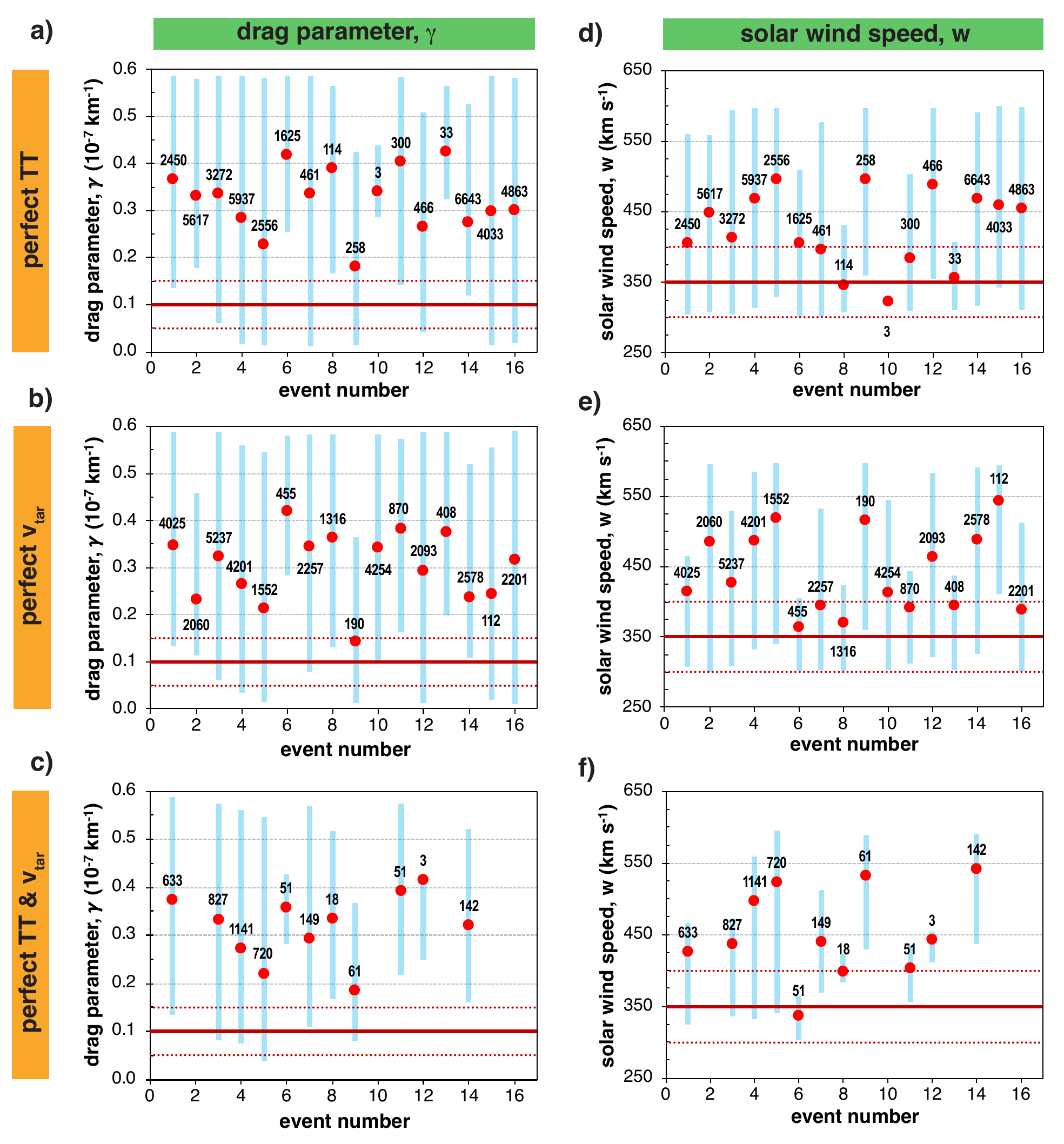}
      \caption{Optimal values for drag parameter, $\gamma$ (left panels: a, b, c) and background solar wind speed, $w$ (right panels: d, e, f) obtained with DBEMv3 to produce: perfect TT within of $\pm$ 1 hour of observed transit time, TT$_{\mathrm{OBS}}$ (upper panels: a, d), perfect arrival speed, $v_{\mathrm{tar}}$ within of $\pm$ 10 km\,s$^{-1}$ of observed arrival speed (middle panels: b, e) and both parameters together: perfect TT and $v_{\mathrm{tar}}$ (lower panels: c, f). The possible range that produced the perfect TT and $v_{\mathrm{tar}}$ values is marked with blue bars and their corresponding median values are shown as the red dots. The number near each red dot indicates the number of found results (values) in altogether 50,000 DBEM runs that matched perfect TT or $v_{\mathrm{tar}}$ or both. The red solid line represent the DBEM input value for $\gamma$ or $w$ with its uncertainty range (dotted red lines) as it was set in \citep{Dumbovic-2018}.}
         \label{Fig8-param-analysis}
   \end{figure}

In the second step, 50,000 DBEMv3 runs were calculated for each event in sample 1 using the obtained constrained range for $\gamma$ and $w$ and other DBEM input parameters shown in Table \ref{table1-CME-list}. From all DBEM runs for each event, the results were filtered to match the perfect TT ($|\Delta$TT$_{\mathrm{err}}| \le 1$ h), perfect $v_{\mathrm{tar}}$ (|$\Delta v_{\mathrm{err}}| \le 10$ km\,s$^{-1}$) or both: perfect TT and $v_{\mathrm{tar}}$ together at the same time. In Figure \ref{Fig8-param-analysis} the obtained values for optimal $\gamma$ (left panels in Figure \ref{Fig8-param-analysis}: a, b, c) and $w$ are presented (right panels in Figure \ref{Fig8-param-analysis}: d, e, f) for each of the analyzed events in the sample 1 (\textit{x-axis}). Almost all events showed the whole range of $\gamma$ and $w$ values that yielded the perfect TT or/and $v_{\mathrm{tar}}$ allowing to calculate the corresponding median values (marked with red dots in Figure \ref{Fig8-param-analysis}) and minimum and maximum values (represented by blue bars). The total number of found values for $\gamma$ or $w$ per event is given as a number near each red dot in Figure \ref{Fig8-param-analysis}.

The optimal drag parameter, $\gamma$ for perfect TT is shown in Figure \ref{Fig8-param-analysis}a and for all analyzed events the optimal median $\gamma$ values (red dots) are much higher than the $\gamma$ value of 0.1 $\times 10^{-7}$ km$^{-1}$ (red solid line) with its uncertainties $\pm 0.05 \times 10^{-7}$ km$^{-1}$ (red dotted lines) used in \citet{Dumbovic-2018} as well as the analysis in Section \ref{DBEM-comparison}. This could be also the reason for the underestimated predicted TT (Table \ref{table1-CME-list}) and $\Delta$TT$_{\mathrm{err}}$ values biased towards negative values (Figure \ref{Fig7-TT-obs-hist}). The optimal $\gamma$ median value for all events is 0.33 $\times 10^{-7}$ km$^{-1}$ (see Table \ref{table3-param-stats}) or about 3 times larger than the previously used value in \citet{Dumbovic-2018} and 5 out of 16 events (31\%) have optimal $\gamma$ range (blue bars) outside the previously used range of 0.1 $\pm$ 0.05 $\times 10^{-7}$ km$^{-1}$. 

\begin{table}
\caption{ Optimal values for all 16 analyzed events together and drag parameter ($\gamma$) as well the background solar wind speed ($w$) calculated with DBEMv3 to produce perfect TT, $v_{\mathrm{tar}}$ and both parameters together: perfect TT and $v_{\mathrm{tar}}$. }
\label{table3-param-stats}                         
\begin{tabular}{l | l | c c c}        
\hline                 
&&perfect TT &perfect $v_{\mathrm{tar}}$&perfect TT \& $v_{\mathrm{tar}}$\\   
\hline                        
& mean&0.32&0.30&0.32\\
drag parameter&median&0.33&0.32&0.33\\
$\gamma$ ($\times 10^{-7}$ km$^{-1}$)&min&0.18&0.14&0.19\\
&max&0.43&0.42&0.41\\
&$\sigma$&0.07&0.07&0.07\\
\hline    
& mean&426&442&453\\
solar wind speed&median&431&421&441\\
$w$ (km\,s$^{-1}$)&min&322&365&338\\
&max&496&544&542\\
&$\sigma$&55&58&64\\ 
\hline                              
\end{tabular}             
\end{table}

The optimal values for the background solar wind speed ($w$) and perfect TT are shown in Figure \ref{Fig8-param-analysis}b in the same way as for $\gamma$. Here is also clear that for more than half of the analyzed events the optimal median $w$ values (red dots) are higher than the previously used ($w$ = 350 km\,s$^{-1}$, $\pm \Delta w$ = 50 km\,s$^{-1}$) in \citet{Dumbovic-2018} and for all events together the median value is 431 km\,s$^{-1}$ (Table \ref{table3-param-stats}). However, the ranges for optimal $w$ are for all events inside the previously used range (300 - 400 km\,s$^{-1}$).

It is interesting to note that for one event (event 10: Feb 23, 2014) just 3 values of optimized $\gamma$ and $w$ could be found with the used DBEM input and constrained range for $\gamma$ and $w$ that were determined in the first step described above. This event also had the largest prediction error in the sample 1, $\Delta$TT$_{\mathrm{err}}$ of almost 33 h (see Table \ref{table1-CME-list}) what could be also the reason why the constrained $\gamma$ and $w$ parameter space could not compensate for such large error to find results. Inspection of the SOHO LASCO CME catalogue \citep{Yashiro-2004, Gopalswamy-2009} revealed that in this period of just a day or two before and after the CME on Feb 23, 2014, there were several strong CMEs that could significantly change the $\gamma$ and $w$ values which are assumed as constant in DBM or that the employed model is not able to fully describe the propagation of this particular CME. For example, \citet{Liu-2014} and \citet{Temmer-Nitta-2015} studied a similar events where preconditioned interplanetary space through the previous CME extremely reduced the drag force with a consequence that the studied CME had a very small deceleration and maintained its high initial speed. Additionally, the reason for large error could also be in unrealistic CME input (\eg CME launch speed, $v_0$). However, detailed analysis of this particular event is out of the scope of this study. 

From Figure \ref{Fig8-param-analysis} it can be also observed that the number of found optimal $\gamma$ and $w$ values (denoted as numbers near each red dot) is directly proportional to the TT prediction errors, $\Delta$TT$_{\mathrm{err}}$ and all events with large number of found optimal values have smaller $\Delta$TT$_{\mathrm{err}}$ in the range of few hours. The average number of optimal values found for perfect TT in the analyzed sample was 2414 per event.

Figure \ref{Fig8-param-analysis}a shows that four CMEs (event 6: Jan 7, 2014, 8: Feb 12, 2014, 11: Feb 25, 2014, 13: Apr 2, 2014) have $\gamma$ median values for perfect TT larger than all for other events ($\gamma > 0.39 \times 10^{-7}$ km$^{-1}$) and their $\gamma$ ranges are significantly higher compared to other events. Furthermore, the mentioned events, compared to the other CMEs, have also significantly lower $w$ median values for perfect TT (Figure \ref{Fig8-param-analysis}d) close to 350 km\,s$^{-1}$ (median for this four events is 370 km\,s$^{-1}$) or within the previously used ranges (300 - 400 km\,s$^{-1}$). The common thing to all four CMEs is the quite large TT prediction error above of 20 h (mean $\Delta$TT$_{\mathrm{err}}$ = -22.98 h) and including event 10 with just three values for optimal $\gamma$ and $w$ found, they have the largest $\Delta$TT$_{\mathrm{err}}$ in the sample 1 (Table \ref{table1-CME-list}). Additionally, all four events are fast CMEs with launch speeds, $v_0$ from 740 to 2399 km\,s$^{-1}$ (mean: 1542 km\,s$^{-1}$). Thus, the reason for such large errors could be connected with the overestimation of CME initial speed that is more pronounced by fast CMEs as already found and discussed in \citet{Dumbovic-2018} and \citet{Mays-2015}.

The optimal $\gamma$ and $w$ for perfect $v_{\mathrm{tar}}$ are shown in Figs. \ref{Fig8-param-analysis}b and \ref{Fig8-param-analysis}e, respectively. Here in comparison to the perfect TT (Figure \ref{Fig8-param-analysis}a), the obtained optimal $\gamma$ median values are very similar, but somewhat lower (median $\gamma$ value for all events is $0.32 \times 10^{-7}$ km$^{-1}$, see Table \ref{table3-param-stats}). The same could be concluded for optimal $w$ and perfect $v_{\mathrm{tar}}$ where the mean $w$ value for all events is 421 km\,s$^{-1}$ (Table \ref{table3-param-stats}). The average number of optimal values found for perfect $v_{\mathrm{tar}}$ in the sample 1 was 2113 per event, similar to the number found for perfect TT (2414 per event).

For 11 out of 16 CMEs (69\%) DBEMv3 was able to find the optimal $\gamma$ and $w$ values for the perfect TT and $v_{\mathrm{tar}}$ simultaneously as shown in Figs. \ref{Fig8-param-analysis}c and f, respectively. The obtained results are similar as for the perfect TT or $v_{\mathrm{tar}}$ where the median value for all events and the optimal $\gamma$ is $0.33 \times 10^{-7}$ km$^{-1}$ and for $w$ is 441 km\,s$^{-1}$ (Table \ref{table3-param-stats}). As expected, in the case of perfect TT and $v_{\mathrm{tar}}$, the number of found optimal values for $\gamma$ and $w$ was significantly smaller (345 per event) due to the fact that both perfect TT and $v_{\mathrm{tar}}$ criteria had to be fulfilled to obtain the optimal $\gamma$ and $w$.

This analysis of the optimal drag parameter, $\gamma$ and solar wind speed, $w$ showed that for most CMEs both $\gamma$ and $w$ values may be underestimated compared to the values used in \citet{Dumbovic-2018}. One of the reasons to explain this issue could be the fact that analyzed CMEs occurred very near the maximum phase of solar cycle. However, the underestimation of $\gamma$ and $w$ should be smaller for the CME samples in the other periods of solar activity and closer to the solar minimum. This also shows the need to evaluate the models with larger CME samples over longer periods to obtain the more general and reliable results.  

\section{DBEMv3 Evaluation with Richardson and Cane (2010) ICME Event List} \label{RC-list-analysis}

As already discussed in chapter \ref{DBEM-errors}, small CME samples during limited periods of time (\eg corresponding only to solar maximum) used for the model evaluation may not provide representative model errors. Thus, it was necessary to employ a larger CME sample over a longer time period covering at least one solar cycle. For that purpose one of the frequently employed lists, the Interplanetary Coronal Mass Ejections (ICME) list compiled by \citet{Richardson-Cane-2010}, R\&C was used\footnote{See \url{http://www.srl.caltech.edu/ACE/ASC/DATA/level3/icmetable2.htm}.}. Initially, R\&C ICME list consists of more than 500 ICMEs, starting in 1996 since the SOHO satellite was launched with LASCO coronographs onboard \citep{Brueckner-1995}. However, this number was reduced to altogether 146 ICMEs\footnote{The complete list of all analyzed events is available as table on \url{https://doi.org/10.6084/m9.figshare.13373867.v1}.} in period from December 1996 to December 2015 as only the ICMEs with all needed DBEM input data available and a clear CME association observed by SOHO/LASCO close to the Sun allowing to determine the longitude of CME source region were selected. Start times of ICMEs from R\&C list was taken as the CME arrival time on Earth and the maximal solar wind speed during the period from the disturbance to the trailing edge of the ICME as in-situ measurement of the CME arrival speed. The DBEM input parameters related to CME properties (CME launch time, projected 2D speed, half-width) were taken from the SOHO/LASCO CME catalogue \citep{Yashiro-2004, Gopalswamy-2009}. Although for Sample 3 the projected 2D CME launch speed ($v_0$) were used, compared to the Sample 1 where 3D speeds were employed and the differences in those speeds can be up to 20\% (\eg \citealt{Jang-2016}), we note that this doesn't introduce a big difference in samples since the other uncertainties related to $v_0$ may be the far larger (\eg estimation of $v_0$ very close to the Sun). Furthermore, the CME launch times were also recalculated to correspond to the starting distance of 20 R$_{\odot}$ (solar radii) for DBEM input. For the CME launch time ($t_0$), half-width ($\lambda$) and longitude ($\phi_{\mathrm{CME}}$) the uncertainties were set as fixed values for all events: $\pm \Delta t_0$ = 30 min, $\pm \Delta \lambda$ = 30$^{\circ}$  and $\pm \Delta \phi_{CME}$ = 10$^{\circ}$. The uncertainty for the CME launch speed ($\pm \Delta v_0$), was estimated as 10\% of the CME launch speed. 

With these input parameters two DBEMv3 evaluations were performed. In the first, the drag parameter, $\gamma$ was set as variable depending on the CME launch speed, $v_0$: for slow CMEs ($v_0 < 600$ km\,s$^{-1}$) $\gamma = 0.5 \times 10^{-7}$ km$^{-1}$, for regular speed CMEs (600 km\,s$^{-1}$ $\ge v_0 \le$ 1000 km\,s$^{-1}$) $\gamma = 0.2 \times 10^{-7}$ km$^{-1}$ and for fast CMEs ($v_0 >$ 1000 km\,s$^{-1}$) $\gamma = 0.1 \times 10^{-7}$ km$^{-1}$ with uncertainties $\pm \Delta \gamma$ of  0.1, 0.075 and 0.05 $\times 10^{-7}$ km$^{-1}$, respectively. This is supported by observations where fast CMEs are usually more massive and have lower $\gamma$ values than in the case of slower CMEs \citep{Vrsnak-2013}.  The background solar wind speed, $w$ was set to $w$ = 450 km\,s$^{-1}$ with uncertainty $\pm \Delta w$ = 50 km\,s$^{-1}$. In the second DBEMv3 evaluation $\gamma$ was taken as fixed value of 0.3 $\times 10^{-7}$ km$^{-1}$ for all events no matter what $v_0$ is, with larger range of uncertainties $\pm \Delta \gamma =  0.2 \times 10^{-7}$ km$^{-1}$. For $w$ was used somewhat lower value of 425 km\,s$^{-1}$ with also larger uncertainties  $\pm \Delta w = $ 100 km\,s$^{-1}$ that corresponds to the mean solar wind speed in OMNI data from 2004 to 2018 \citep{Amerstorfer-2021}. The second evaluation was based on $\gamma$ and $w$ analysis obtained in Section \ref{DBEM-gamma-w} where it was found that for the majority of analyzed events to get perfect TT, the optimal $\gamma$ was around $0.3 \times 10^{-7}$ km$^{-1}$. All DBEMv3 input parameters with corresponding uncertainties for two described evaluations (variable and fixed $\gamma$) are summarized in Table \ref{table4-DBEM-input}. For the both evaluations and each event the 50,000 DBEM runs were performed. 

\begin{table}
\caption{ Summary of all DBEMv3 input parameters and uncertainties for evaluation with the variable drag parameter $\gamma$ depending on $v_0$ and evaluation with fixed $\gamma = 0.3 \times 10^{-7}$ km$^{-1}$. R\&C list denotes that input parameter was taken from Richardson\&Cane ICME list. $w$ is background solar wind speed, $t_0$ CME launch date and time, $v_0$ CME launch speed, $\lambda$ CME’s angular half-width, $\phi_{\mathrm{CME}}$ longitude of the CME source region and $R_0$ CME starting radial distance. }
\label{table4-DBEM-input}                         
\begin{tabular}{l | c c | c c }        
\hline
& \multicolumn{2}{c}{variable $\gamma$} & \multicolumn{2}{|c}{fixed $\gamma$} \\  
parameter&input&$\pm$uncertainty&input&$\pm$uncertainty\\    
\hline                        
$\gamma$&0.5 \tiny{($v_0 < 600$ km\,s$^{-1}$)}&0.1& &\\
\tiny{($\times 10^{-7}$ km$^{-1}$)}&0.2 \tiny{(600 $\ge v_0 \le$ 1000 km\,s$^{-1}$)}&0.075&0.3&0.2\\
&0.1 \tiny{($v_0 >$ 1000 km\,s$^{-1}$)}&0.05& &\\
\hline
$w$ \tiny{(km\,s$^{-1}$})&450&50&425&100\\
$t_0$&R\&C list&30 min&R\&C list&30 min\\
$v_0$ \tiny{(km\,s$^{-1}$)}&R\&C list&10\%&R\&C list&10\%\\
$\lambda$  \tiny{($^{\circ}$)}&R\&C list&30&R\&C list&30\\
$\phi_{\mathrm{CME}}$ \tiny{($^{\circ}$)}&R\&C list&10&R\&C list&10\\
$R_0$ \tiny{(in R$_{\odot}$)}&20&-&20&-\\
\hline                              
\end{tabular}             
\end{table}

\begin{figure}
   \centering
   \includegraphics[width=\textwidth]{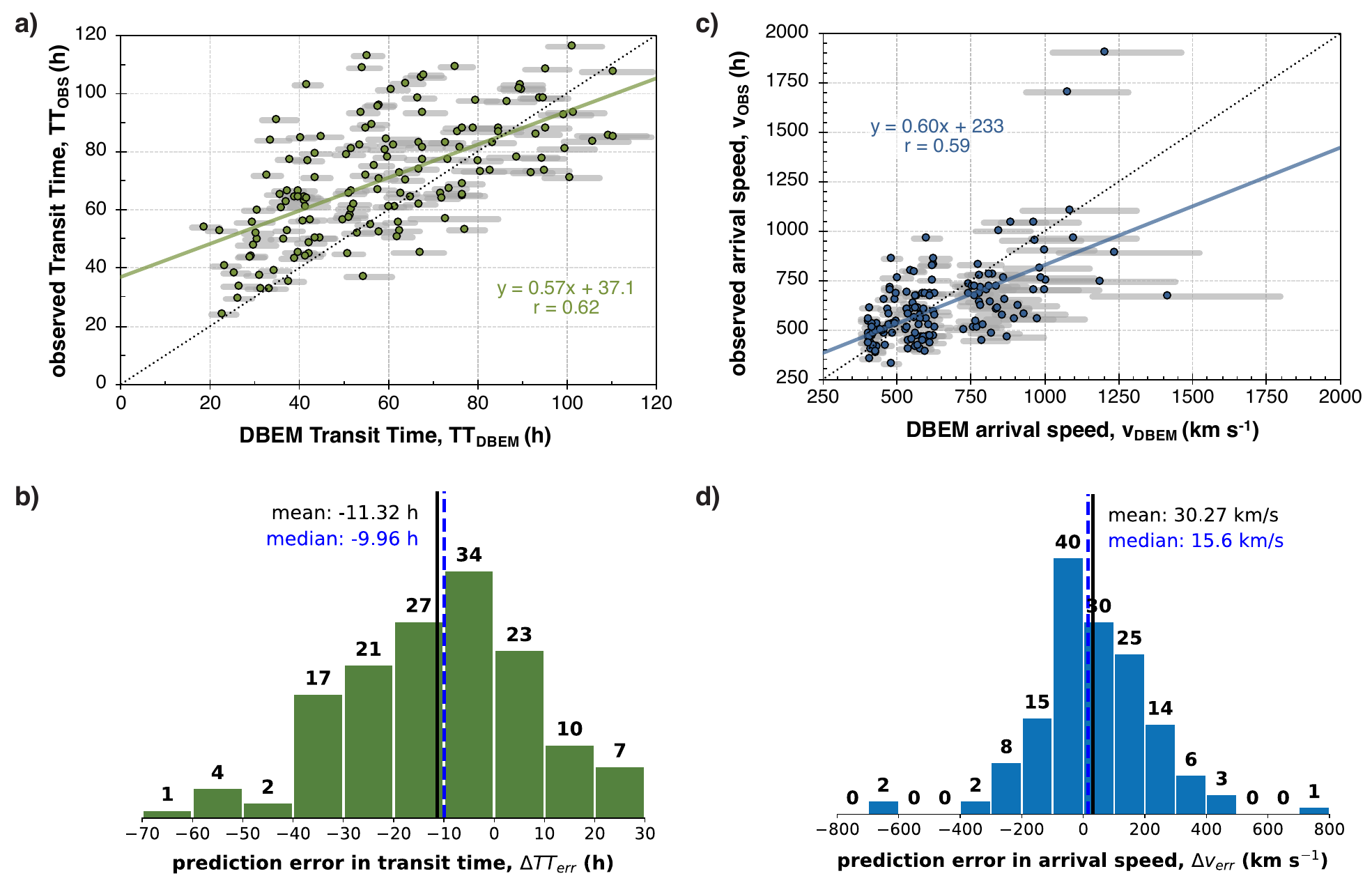}
      \caption{Comparison between the observed and DBEM predicted: a) transit time, TT expressed in hours and c) CME arrival speed, $v_{\mathrm{tar}}$ in km\,s$^{-1}$ (upper panels). The black dotted line represent the identity line (perfect match). The green (TT) and blue ($v_{\mathrm{tar}}$) solid lines show the best linear fit determined by shown equation with the correlation coefficient. The grey bars around the points are 95\% confidence intervals calculated with DBEM. Lower panels show the corresponding histograms with prediction errors for: b) transit time, $\Delta$TT$_{\mathrm{err}}$ expressed in hours and d) arrival speed, $\Delta v_{\mathrm{err}}$ in km\,s$^{-1}$. On top of each bin is shown the count of events in each bin. The shown results are for the DBEM predictions with the variable $\gamma$ values ($0.1 - 0.5 \times 10^{-7}$ km$^{-1}$) and $w$ = 450 km\,s$^{-1}$.}
         \label{Fig9-obs-dbem}
   \end{figure}

The results of the first DBEMv3 evaluation with variable $\gamma$ are presented in Figure \ref{Fig9-obs-dbem}. The observed versus predicted DBEM TT (Figure \ref{Fig9-obs-dbem}a) show similar results that were found already with the smaller CME sample in Section \ref{DBEM-comparison} (Figure \ref{Fig6-obs-DBEM}). The predicted TT is underestimated for fast CMEs with shorter transit times and the TT is overestimated for slower CMEs with longer transit times. The best linear fit indicates a significantly better correlation coefficient ($r = 0.62$) than in the case of the CME sample 1 and 2 in Section \ref{DBEM-comparison} ($r = 0.49$). Only 21 of 146 analyzed ICMEs (14.4\%) had observed TT$_{\mathrm{OBS}}$ inside the DBEM predicted uncertainties range for TT (95\% confidence intervals, denoted by grey bars in Figure \ref{Fig9-obs-dbem}). The corresponding histogram of prediction error in transit time, $\Delta$TT$_{\mathrm{err}}$ (Figure \ref{Fig9-obs-dbem}b) clearly indicates the shift towards the negative prediction errors (underestimated TT) where the mean error (ME) is $-11.32$ h. The mean absolute error (MAE) for $\Delta$TT$_{\mathrm{err}}$ is 17.26 h and the root mean square error (RMSE) is 21.81 h (all prediction errors are listed in Table \ref{table5-RC-lst-errors}). The larger sample 3 may be the reason for significantly larger errors that were obtained than in the case of evaluation with the smaller limited CME sample (Table \ref{table2-DBEM-stats}). In the same way, the observed and predicted CME arrival speed, $v_{\mathrm{tar}}$ are plotted in Figure \ref{Fig9-obs-dbem}c. Here the predicted $v_{\mathrm{tar}}$ are significantly overestimated for fast CMEs with some underestimation of predicted $v_{\mathrm{tar}}$ for slow CMEs. This is also revealed in the presented histogram for $\Delta v_{\mathrm{err}}$ prediction error (Figure \ref{Fig9-obs-dbem}d) showing a small shift towards positive values of $\Delta v_{\mathrm{err}}$ and ME = 30 km\,s$^{-1}$ (MAE = 139 km\,s$^{-1}$, RMSE = 190 km\,s$^{-1}$, see Table \ref{table5-RC-lst-errors}). DBEM 95\% confidence intervals for $v_{\mathrm{tar}}$ were in 17.8\% cases (26 of 146 events) within the observed arrival speed, $v_{\mathrm{OBS}}$ what is similar as in the case of TT. It should be noted that two ICMEs with very high observed arrival speed, $v_{\mathrm{OBS}}$ larger than 1700 km\,s$^{-1}$ are outliers from all other analyzed ICMEs. Both events (Oct 28, 2003 11:10 UTC and Oct 29, 2003 20:49 UTC) were among the strongest ICMEs in the last solar cycle and part of Haloween stroms in 2003 with CME launch speeds greater than 2000 km\,s$^{-1}$ that produced major geomagnetic storms on the Earth. For these two events the DBEM obtained two largest $\Delta v_{\mathrm{err}}$ prediction errors of $-695$ and $-620$ km\,s$^{-1}$. This actually shows that for such extreme CMEs, probably special caution is needed to fine tune the DBEM input parameters such as $\gamma$ and $w$ (in this case smaller $\gamma$ and larger value for $w$) due to highly disturbed interplanetary space and the specific CME properties. 

\begin{table}
\caption{ Various prediction errors in transit time, $\Delta$TT$_{\mathrm{err}}$ (hours) and in arrival speed, $\Delta v_{\mathrm{err}}$ (km\,s$^{-1}$) for evaluation with the variable drag parameter $\gamma$ depending on $v_0$ and fixed $\gamma = 0.3 \times 10^{-7}$ km$^{-1}$. ME is mean error, MAE is mean absolute error, RMSE is root mean square error and $\sigma$ is calculated standard deviation. }
\label{table5-RC-lst-errors}                         
\begin{tabular}{l | c c | c c }        
\hline
& \multicolumn{2}{c}{$\Delta$TT$_{\mathrm{err}}$ (h)} & \multicolumn{2}{c}{$\Delta v_{\mathrm{err}}$ (km\,s$^{-1}$)} \\  
error & variable $\gamma$ & fixed $\gamma$ & variable $\gamma$ & fixed $\gamma$  \\    
\hline                        
ME&-11.32&-3.90&30&-102\\
MAE&17.26&14.54&139&135\\
RMSE&21.81&18.57&190&208\\
median&-9.95&-3.21&16&-76\\
min&-61.14&-52.57&-695&-1258\\
max	&29.85&41.31&746&147\\
$\sigma$&18.70&18.22&188&182\\
\hline                              
\end{tabular}             
\end{table}

\begin{figure}
   \centering
   \includegraphics[width=\textwidth]{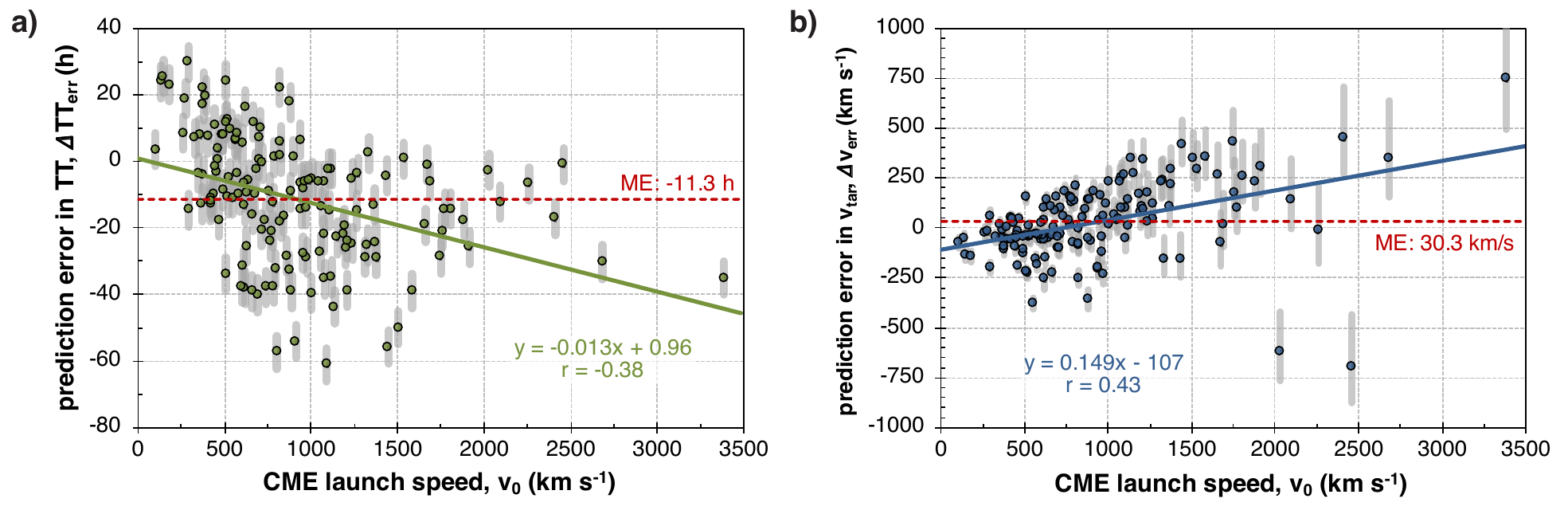}
      \caption{Prediction error in a) transit time, TT$_{\mathrm{err}}$ expressed in hours and b) arrival speed, $v_{\mathrm{err}}$ plotted against the CME input speed, $v_0$ expressed in km\,s$^{-1}$. The red dashed line denotes mean error (ME) and green (TT$_{\mathrm{err}}$) and blue ($v_{\mathrm{tar}}$) solid lines show the best linear fit determined by shown equation and the correlation coefficient. 95\% confidence intervals calculated with DBEM are shown with grey bars around the points. The shown prediction errors are for the DBEM calculations performed with the variable $\gamma$ values ($0.1 - 0.5 \times 10^{-7}$ km$^{-1}$) and $w = 450$ km\,s$^{-1}$.}
       \label{Fig10-dTT-dvtar}
\end{figure}

It was already reported by \citet{Mays-2015} and \citet{Dumbovic-2018} that both ENLIL and DBEM may predict fast CMEs to arrive earlier than they were observed, resulting in negative prediction errors, $\Delta$TT$_{\mathrm{err}}$. The dependence of prediction error $\Delta$TT$_{\mathrm{err}}$ on the CME launch speed, $v_0$ is shown in Figure \ref{Fig10-dTT-dvtar}a. From the best linear fit (green solid line) with correlation cofficient $r = -0.38$ it can be seen that almost all fast CMEs above $\approx$ 1000 km\,s$^{-1}$ have negative values of $\Delta$TT$_{\mathrm{err}}$ (underestimated TT). On the other hand, CMEs slower than $\approx$ 1000 km\,s$^{-1}$ have the tendency that TT is overestimated. One of the reason for underestimated predicted TT for fast CMEs could be due to the overestimation of the CME launch speed, $v_0$. Another reason could be related to model limitations itself and the fast CME rapid deceleration phase \citep{Liu-2013, Liu-2016} that is not taken into account by DBM. The prediction error in the arrival speed, $\Delta v_{\mathrm{err}}$ shows also a notable dependence on $v_0$ as shown in Figure \ref{Fig10-dTT-dvtar}b. Here, the arrival speed, $v_{\mathrm{tar}}$ for fast CMEs has the tendency to be overestimated what is supported by the best linear fit (blue solid line, $r = 0.43$). It should be noted again that  two outliers with large negative $\Delta v_{\mathrm{err}}$ are the Halloween storms in 2003. 

\begin{figure}
   \centering
   \includegraphics[width=\textwidth]{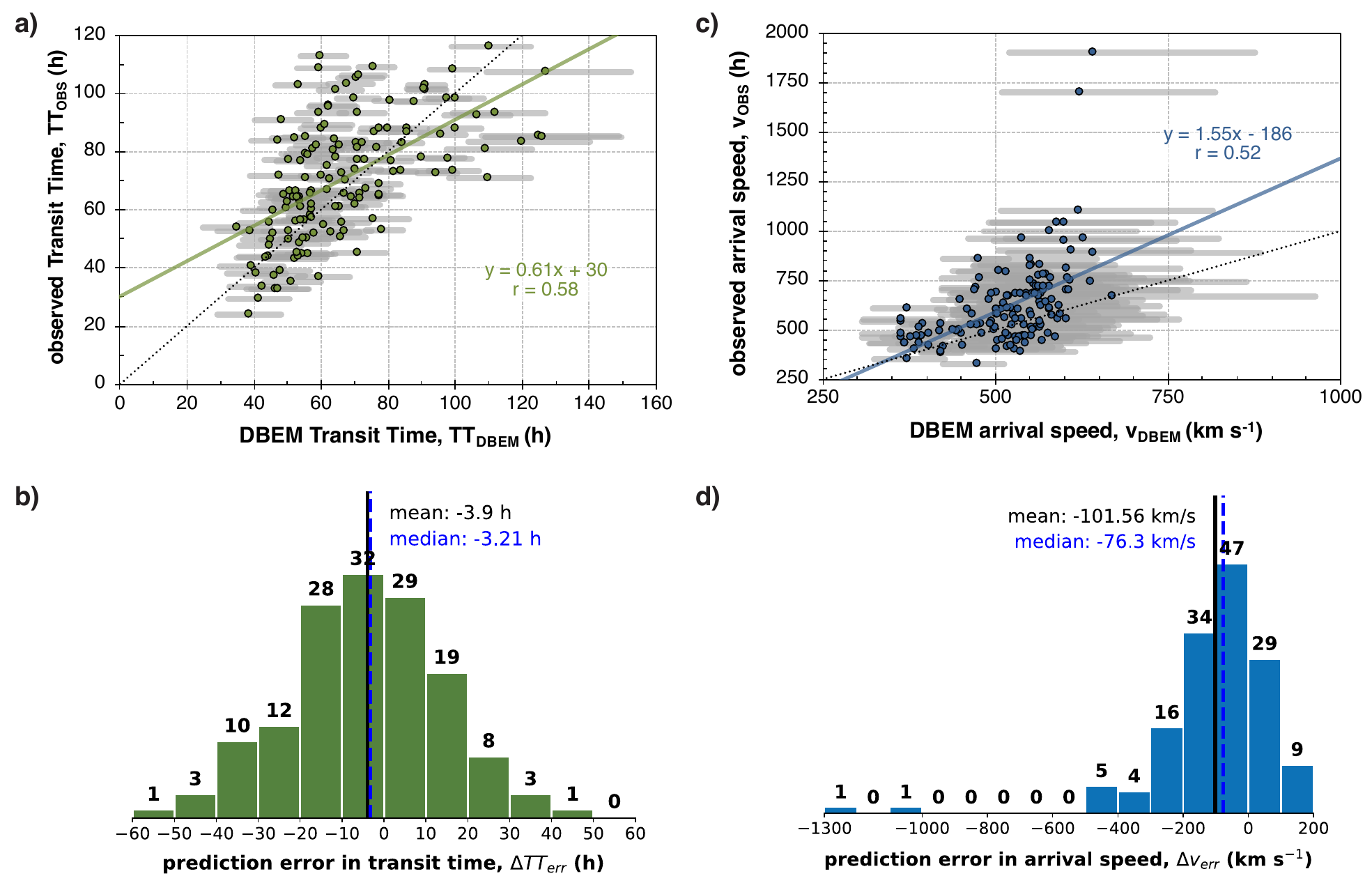}
      \caption{Comparison between the observed and DBEM predicted: a) transit time, TT expressed in hours and c) CME arrival speed, $v_{\mathrm{tar}}$ expressed in km\,s$^{-1}$ (upper panels). The black dotted line represent the identity line (perfect match). The green (TT) and blue ($v_{\mathrm{tar}}$) solid lines show the best linear fit determined by shown equation with the correlation coefficient. The grey bars around the points are 95\% confidence intervals calculated with DBEM. Lower panels show the corresponding histograms with prediction errors for: b) transit time, $\Delta$TT$_{\mathrm{err}}$ expressed in hours and d) arrival speed, $\Delta v_{\mathrm{err}}$ expressed in km\,s$^{-1}$. On top of each bin is shown the count of events in each bin. The shown results are for the DBEM predictions with the fixed $\gamma$ value ($0.3 \times 10^{-7}$ km$^{-1}$) and $w$ = 425 km\,s$^{-1}$.}
       \label{Fig11-obs-dbem}
\end{figure}

The results of the second DBEMv3 evaluation with fixed $\gamma = 0.3 \times 10^{-7}$ km$^{-1}$ are presented in Figure \ref{Fig11-obs-dbem} in the same way as for the first evaluation. Surprisingly, the second evaluation resulted generally in significantly smaller prediction errors $\Delta$TT$_{\mathrm{err}}$ and ME of $-3.9$ h, \ie almost three times smaller compared to the first evaluation. The obtained MAE was 14.54 h and RMSE  18.57 h (see Table \ref{table5-RC-lst-errors}). The larger input uncertainties for $\gamma$ and $w$ produced also the larger DBEM prediction errors for TT and in 28.8\% cases for all analyzed ICMEs the observed TT$_{\mathrm{OBS}}$ were found within the TT prediction uncertainties which is twice larger than the value found in the previous evaluation with variable $\gamma$. The corresponding histogram showing the TT prediction error (Figure \ref{Fig11-obs-dbem}b) also reveals a smaller bias towards negative $\Delta$TT$_{\mathrm{err}}$ or underestimated predicted TT$_{\mathrm{DBEM}}$. However, despite the fact that overall prediction error statistics with fixed $\gamma$ are better than in the case of variable $\gamma$, in some cases the predictions get worse (in particular for fast CMEs). This is also reflected by a lower correlation coefficient ($r = 0.58$) of the best linear fit in Figure \ref{Fig11-obs-dbem}a. An additional drawback of using fixed $\gamma$ is the much larger prediction error for arrival speeds, $\Delta v_{\mathrm{err}}$ shown in Figure \ref{Fig11-obs-dbem}c and visible also in related histogram (Figure \ref{Fig11-obs-dbem}d). In overall statistic, $\Delta v_{\mathrm{err}}$ obtained with fixed $\gamma$ has several times larger ME of $-102$ km\,s$^{-1}$, MAE 135 km\,s$^{-1}$ and RMSE 208 km\,s$^{-1}$ compared to evaluation with variable $\gamma$ (Table \ref{table5-RC-lst-errors}). 37\% of the analyzed ICMEs had observed $v_{\mathrm{tar}}$ within the DBEM predicted $v_{\mathrm{tar}}$ errors what is double value of the ratio in the evaluation with variable $\gamma$. However, the reason for the higher ratio concerning prediction errors was also due to the larger $\gamma$ and $w$ uncertainties employed than in the case of the evaluation with variable $\gamma$. 

\begin{figure}
   \centering
   \includegraphics[width=\textwidth]{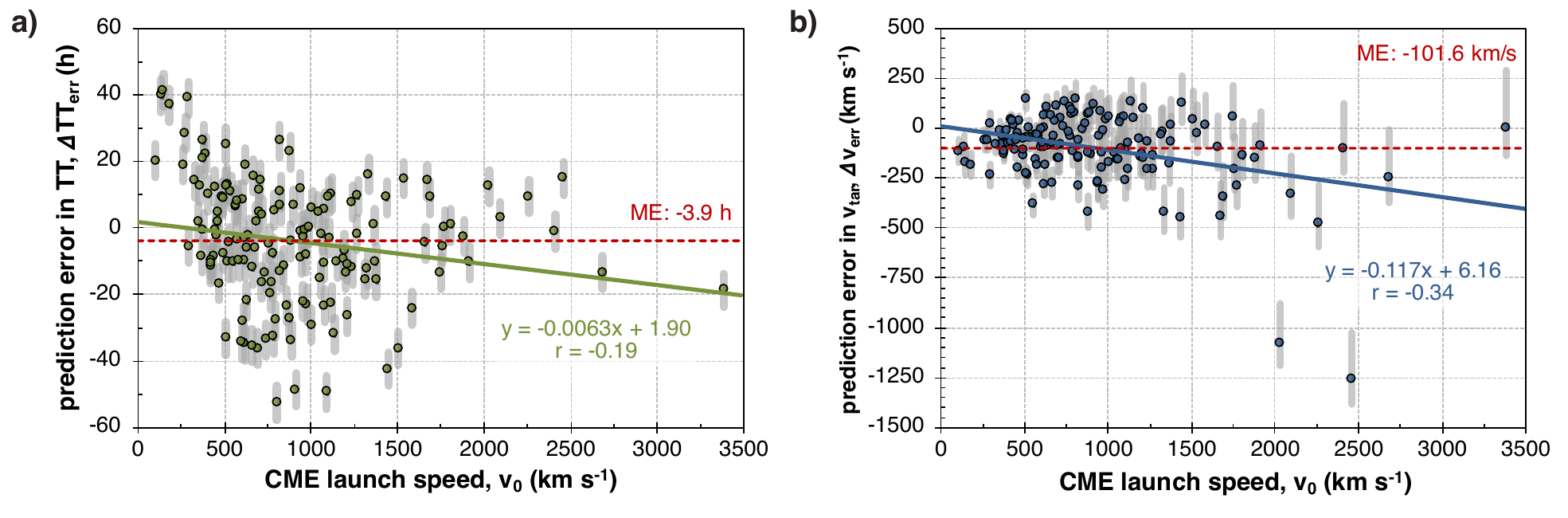}
      \caption{Prediction error in a) transit time, $\Delta$TT$_{\mathrm{err}}$ expressed in hours and b) arrival speed, $\Delta v_{\mathrm{err}}$ plotted against the CME input speed, $v_0$ expressed in km\,s$^{-1}$. The red dashed line denotes mean error (ME) and green (TT) and blue ($v_{\mathrm{tar}}$) solid lines show the best linear fit determined by shown equation and the correlation coefficient. 95\% confidence intervals calculated with DBEM are shown with grey bars around the points. The shown prediction errors are for the DBEM calculations performed with the fixed $\gamma$ value ($0.3 \times 10^{-7}$ km$^{-1}$) and $w = 425$ km\,s$^{-1}$.}
       \label{Fig12-dTT-dvtar}
\end{figure}

The dependence of prediction errors $\Delta$TT$_{\mathrm{err}}$ and $\Delta v_{\mathrm{err}}$ on the CME launch speed, $v_0$ was also analyzed for the fixed $\gamma$ evaluation and it is presented in Figure \ref{Fig12-dTT-dvtar}. Interestingly, the dependence of $\Delta$TT$_{\mathrm{err}}$ on $v_0$ is somewhat weaker which is supported by the lower linear fit correlation coefficient of $r=-0.19$ and the slope of the regression line is flatter indicating that the bias towards negative $\Delta$TT$_{\mathrm{err}}$ for larger $v_0$ is reduced (Figure \ref{Fig12-dTT-dvtar}a). Compared to the first evaluation, for $\Delta v_{\mathrm{err}}$ the linear fit slope changes direction from positive to negative value and $v_{arr}$ has now tendency to be underestimated for increasing $v_0$ (Figure \ref{Fig12-dTT-dvtar}b). $\Delta v_{\mathrm{err}}$ dependence on $v_0$ is also characterized by slightly lower correlation coefficient of $r=-0.34$ than it was the case in the first evaluation ($r=0.43$). Here again, the two Haloween storms in 2003 are very strong negative outliers in $\Delta v_{\mathrm{err}}$.

\section{Summary and Conclusions}

We evaluate and describe in detail for the first time the most recent version of the Drag-based Ensemble Model, DBEMv3 that provides probabilistic predictions of the coronal mass ejections (CMEs) propagation in ecliptic plane for the Earth or any object in the solar system. It determines the probability that the CME hits the target, the most likely CME arrival time and arrival speed together with a quantification of the prediction uncertainties and confidence intervals. The recently developed DBEMv3 incorporates various improvements and updates from DBEMv1 and DBEMv2 versions. It includes now also the vizualization of CME propagation (CME geometry and kinematic plots) and Graduated Cylindrical Shell (GCS) option for the CME geometry input. Since DBEM has the advantage to be a very fast, reliable and simple model, it is suited for fast real-time space-weather forecasting. The online DBEMv3 web application is already available as one of the European Space Agency (ESA) space situational awareness (SSA) services\footnote{See \url{http://swe.ssa.esa.int/heliospheric-weather}.}. 

A comparison of DBEMv3 with the previous DBEMv1 version shows the excellent agreement for all calculated output parameters with the correlation coefficient $r > 0.99$. The evaluation with the same CME-ICME sample as in \citet{Dumbovic-2018} (sample 1) showed slight improvement in DBEMv3 prediction errors compared to DBEMv1. However, these differences are rather small (\eg half hour for transit time, TT) and few orders of magnitude smaller than the "standard" drag-based model (DBM) and the effects of CME input errors. 

The drag parameter ($\gamma$) and the background solar wind speed ($w$) are two important DBEM input parameters, very much related to the DBEM prediction errors and very difficult to asses or to measure directly. With that in mind, the reverse modelling with DBEMv3 using  sample 1 was performed to find optimal $\gamma$ and $w$ parameters to match the observed TT and CME arrival speed at target, $v_{\mathrm{tar}}$. In general, for most events, the found optimal $\gamma$ was three times higher ($\gamma = 0.3 \times 10^{-7}$ km$^{-1}$) than previously used in the \citet{Dumbovic-2018} evaluation ($\gamma = 0.1 \times 10^{-7}$ km$^{-1}$) and the optimal solar wind speed ($w$) was around 430 km\,s$^{-1}$ which is also higher than the previously used ($w$ = 350 km\,s$^{-1}$). One of the possible explanations for a higher value of $w$ could be due to the numerous CMEs in solar maximum affecting $w$, where $w$ can be increased up to 30\% about 3 to 6 days after the start of the CME \citep{Temmer-2017}.

Since it was also noted in this analysis that smaller CME samples with only ten to twenty events during a limited period of time corresponding only to certain solar activity phase may not provide representative model errors, the ICME list compiled by \citet{Richardson-Cane-2010} was used. The DBEMv3 evaluation with altogether 146 selected CME-ICME pairs (sample 3) during the period of almost two solar cycles (December 1996 - December 2015) showed slightly larger (also due to the size of analyzed sample), but comparable prediction errors to the previous DBEM \citep{Dumbovic-2018} and other model evaluations like ENLIL \citep{Mays-2015}. For example, the prediction errors for TT and variable $\gamma$ model input (dependent on CME launch speed, $\gamma = (0.1 - 0.5) \times 10^{-7}$ km$^{-1}$) was found to have mean error (ME) of $-11.32$ h, mean absolute error (MAE) of 17.26 h and root mean square error of 21.81 h. For fast CMEs there is a clear bias towards negative prediction errors for TT (the predicted ICME arrival time is too early compared to observed ICME arrival). An explanation for that systematic error could be the overestimation of the CME launch speed or the physical limitations in DBM model itself. It should be also noticed that for several extremely fast and strong ICMEs (Halloween 2003 events) the largest prediction errors were obtained and thus such events would require special fine tuning of DBEM input parameters due to the complex heliospheric conditions. 

Based on the results obtained for optimal $\gamma$ and $w$ parameters, the additional DBEMv3 evaluation was performed with 146 ICMEs and fixed $\gamma$ value ($\gamma = 0.3 \times 10^{-7}$ km$^{-1}$). Interestingly, the employed $\gamma$ with higher value improved in overall the TT prediction significantly (ME = $-3.9$ h, MAE = 14.54 h, RMSE = 18.57 h) and reduced the bias towards the negative TT prediction errors for fast CMEs. At the same time prediction errors for $v_{\mathrm{tar}}$ were increased several times (ME = $-102$ km\,s$^{-1}$ compared to ME = $-30$ km\,s$^{-1}$). This again shows the importance of drag parameter $\gamma$ that was possibly underestimated in previous studies (\eg in \citealp{Dumbovic-2018}). Furthermore, due to the larger input uncertainties for $\gamma$ and $w$ employed in evaluation with fixed $\gamma$, the larger DBEM prediction uncertainties were obtained and almost one third (28.8\%) of the analyzed ICMEs observed TT$_{\mathrm{OBS}}$ were within predicted TT uncertainties which was two times larger compared to the evaluation with variable $\gamma$ (14.4\%). Although, the even bigger input uncertainties would increase the ratio of TT$_{\mathrm{OBS}}$ within predicted TT uncertainties, ithe much larger DBEM prediction uncertainties in general wouldn't bring some benefits to CME propagation forecasting. 

The two different evaluations of DBEMv3 performed here with sample 3 (R\&C ICME list) show the importance of $\gamma$ and $w$ which are usually hard to select or estimate due to the lack of measurements to reliably provide these parameters. Although, it was shown here that a fixed $\gamma$ value for all events in general may be the simpler choice and may significantly reduce the model errors, there should be caution due to the fact that single CME prediction errors (\eg fast CMEs that are important for space weather forecast) may be in some cases larger than when using $\gamma$ dependent on the CME launch speed, $v_0$. On the other hand, some very fast and strong CMEs, like two Halloween storms in 2003 that are clearly outliers in the both evaluations, should be given special attention and $\gamma$ and $w$ parameters should be carefully tuned to obtain reliable model predictions. However, detailed analysis for these two events is out of the scope of this study. 

Since DBM analytical solution assumes the constant drag parameter $\gamma$ and background solar wind conditions, it is expected that DBEM does not give the best CME arrival time predictions in a complex heliospheric environment or \eg during solar maximum \citep{Vrsnak-2013}. During that time due to a high number of CMEs a numerous CME-CME interaction events can happen \citep{Liu-2014, Temmer-2012, Temmer-2017, Rodriguez-2020} or CMEs may pass trough the high speed streams \citep{Vrsnak-2010} thus in both cases significantly influencing the value of the $\gamma$ and $w$ input parameters. Therefore, the analysis of such events and development the methods to change both $\gamma$ and $w$ parameters along the CME propagation path to improve the forecast will be the topic of the future research and DBEM improvement.

%

%
\begin{acks}
We acknowledge funding from the EU H2020 grant agreement No. 824135 (SOLARNET) and support by the Croatian Science Foundation under the project 7549 (MSOC). M.D. and B.V. acknowledge support by the Croatian Science Foundation under the project IP-2020-02-9893 (ICOHOSS).
\end{acks}

\section*{Disclosure of Potential Conflicts of Interest}
The authors declare that they have no conflicts of interest.

%
%
\bibliographystyle{spr-mp-sola}
\bibliography{references} 
%
%
%
%

\end{article} 
\end{document}